\documentclass[aps,twocolumn]{revtex4}
\usepackage{color}
\usepackage{psfrag}
\usepackage{dcolumn}
\usepackage{bm}
\usepackage{float}
\usepackage[latin1]{inputenc}
\usepackage[spanish,english]{babel}
\usepackage{amsfonts}
\usepackage{amssymb,epsf}
\usepackage{graphicx}
\usepackage{slashed}
\usepackage{epstopdf}
\usepackage{amsmath,amssymb}
\usepackage{pdflscape}
\usepackage{adjustbox}
\usepackage{hyperref}
\usepackage{subcaption}
\usepackage{appendix}
\begin{document}

	\author{J. Sedaghat\footnote{
			email address: J.sedaghat@shirazu.ac.ir}, G. H. Bordbar\footnote{
			email address: ghbordbar@shirazu.ac.ir (corresponding author)}, M. Haghighat\footnote{
			email address: m.haghighat@shirazu.ac.ir }, S. M. Zebarjad\footnote{
			email address: zebarjad@shirazu.ac.ir}}

	\affiliation{Physics Department and Biruni Observatory, Shiraz University, Shiraz 71454, Iran}
	\title{Two--fluid CFL strange quark stars with scalar dark matter: critical mass and mass--gap implications}
	\begin{abstract}
	{We investigate the structure of strange quark stars (SQSs) in the color--flavor--locked (CFL) phase in the presence of scalar bosonic dark matter within a two--fluid formalism. The quark matter equation of state (EOS) is obtained from perturbative QCD, incorporating the latest Particle Data Group values for the running coupling and strange quark mass, while the dark matter component is modeled as a self-interacting Bose--Einstein condensate. By considering different dark matter masses and varying the pairing gap $\Delta$ and {the central dark matter pressure fraction} $f_r$, we analyze the impact of dark matter on the structural properties of SQSs, including the maximum gravitational mass $M_{\mathrm{TOV}}$, the ratio of dark matter to strange-quark-matter radii $R_{\mathrm{DM}}/R_{\mathrm{SQM}}$, and the dimensionless tidal deformability $\Lambda$. {We further examine the compatibility of the resulting mass--radius relations with the recent NICER measurements of compact stars.} Within the parameter space considered in this study, we find that $M_{\mathrm{TOV}}$ exhibits a non-monotonic dependence on the dark matter mass, with a critical value beyond which $M_{\mathrm{TOV}}$ decreases. We also show that some pure CFL strange quark star configurations, particularly those associated with very stiff EOSs and larger maximum masses, may not simultaneously remain compatible with the $\Lambda$ range inferred from GW170817 while occupying the lower mass--gap region. In contrast, the inclusion of dark matter allows two-fluid CFL strange quark star configurations to reproduce the observed properties of massive compact objects in the lower mass--gap region, such as the secondary component of GW190814, while remaining qualitatively compatible with the $\Lambda$ range inferred from GW170817. We note, however, that the GW170817 constraints were originally inferred within single-fluid compact-star frameworks and therefore provide only {qualitative guidance} for the present two-fluid halo configurations. Our results suggest that exotic compact-star configurations may populate part of the conventionally defined lower mass--gap region.}

	\end{abstract}
	
	\maketitle
	
	\section{Introduction}
	The existence of dark matter remains one of the greatest unresolved mysteries in modern physics. It accounts for approximately 27\% of the universe's total mass-energy content, yet it does not interact with electromagnetic radiation, making it invisible to direct observations \cite{Martino2020,Planck Collaboration 2014,Persic1996,Bertone2018}. While the Standard Model of particle physics successfully describes known fundamental particles and their interactions, it lacks a viable dark matter candidate. None of its particles possess the necessary characteristics\textemdash such as long-term stability, non-relativistic behavior, and absence of electromagnetic interaction\textemdash to explain astrophysical and cosmological evidence \cite{Bertone2005}. A range of indirect observations, including galaxy rotation curves, gravitational lensing, and anisotropies in the cosmic microwave background, strongly support the presence of dark matter, yet its fundamental nature remains unknown \cite{Zwicky 1933, Allen2011, Frenk2012, Bennett2013}. Extensions to the Standard Model, such as theories involving Weakly Interacting Massive Particles (WIMPs) \cite{Weinberg1977,Goodman1985,Vysotskii1977,Adhikari2023}, axions, or sterile neutrinos, aim to address this gap, though no conclusive experimental detection has been made to date \cite{Weinberg1978,Preskill1983,Abbott1983}. Current searches\textemdash through direct detection experiments, high-energy collider studies, and astrophysical observations\textemdash continue to seek insights into the true nature of dark matter \cite{Roszkowski2018}.
	
	The inclusion of dark matter in the modeling of compact stars offers an intriguing avenue for exploring its astrophysical effects. Compact stars\textemdash remnants of massive stellar evolution\textemdash include neutron stars \cite{Wiringa, Stone2007, Pearson,Negreiros,BordbarNS1,BordbarNS2,BordbarNS3,BordbarNS4}, hybrid stars 	\cite{Blaschke2001,Burgio2003,Alford2005,Pal2023,Rather2023,Li2023,BordbarNSQ1,BordbarNSQ2}, and quark stars \cite{Michel1988,Drago2001,Kurkela2010,Wang2019,Deb2021,Sedaghat2021,JSedaghat,sedaghat,sedaghatannals,sedaghatonefluid, Shahrbaf2025}, whose internal structures are governed by their respective equations of state (EOSs). The presence of dark matter can modify these EOSs, thereby modifying the macroscopic properties of the stars. Dark matter within compact stars is typically modeled using two main approaches: the two--fluid model and the one--fluid model. In the two--fluid model, dark matter and baryonic matter are treated as separate components with independent EOSs, coupled only through gravity. This framework is particularly suited for configurations where dark matter forms a halo or core, influencing global stellar properties such as mass, radius, and dimensionless tidal deformability $\Lambda$ \cite{Chaudhuri2024,Anzuini,Routaray,Das2019,Panotopoulos2017,Karkevandi2022,Karkevandi2024,Karkevandi2024b,Yang2025}. In contrast, the one--fluid model assumes direct interactions between dark matter and baryonic matter\textemdash often motivated by physics beyond the Standard Model\textemdash which lead to a unified EOS  \cite{sedaghatonefluid,Chaudhuri2024,Shahrbaf2025}.
	
	Incorporating dark matter into compact star models may also help resolve outstanding questions about massive compact objects, particularly those that appear within the so-called mass--gap. 
	The existence of a mass--gap in the population of compact objects, typically in the range of $2.5$--$5\,M_{\odot}$, has long been an open question in astrophysics \cite{Bailyn,ApJ725,Belczynski}. Recent observations of compact objects in this interval indicate that the interpretation of a strict mass--gap may require revision.
	 For instance, the secondary component of GW190814, with a mass of approximately 2.6 $M_{\odot}$ \cite{Abbott896}, remains puzzling\textemdash it could represent either the most massive neutron star ever observed or the least massive black hole \cite{Koliogiannis}. Similarly, PSR J0952-0607, with a mass of 2.35 $\pm$ 0.17 $M_{\odot}$, stands as the most massive known neutron star, lying near the lower edge of the mass--gap \cite{J0952}, while PSR J0740+6620, with a mass of 2.08 $\pm$ 0.07 $M_{\odot}$ \cite{Miller918}, further suggests that the upper mass limit for neutron stars may extend into this range. These findings indicate that the mass--gap is not entirely empty of compact objects, and that its boundaries may require revision. 
	Furthermore, discovery of massive compact objects may offer new insights into the composition of these stars.  For example, possible  detections of compact objects in the 2.5--3 $M_{\odot}$ range, may show the need for hybrid stars with quark cores, pure quark stars \cite{Annala}, or even the need for inclusion of dark matter in compact star \cite{Chaudhuri2024,Anzuini,Routaray,Das2019,Panotopoulos2017,Karkevandi2022,Karkevandi2024,Karkevandi2024b,Yang2025}.
	
{In this paper, we investigate quark stars composed of strange quark matter (SQM) in the color--flavor--locked (CFL) phase admixed with dark matter within the framework of the two--fluid model. Our goal is to explore whether such objects can populate the lower mass--gap region while remaining {qualitatively compatible} with the $\Lambda$ range inferred from GW170817. We also compare these results with pure CFL strange quark star (SQS) configurations (without dark matter), which for some parameter choices associated with very stiff EOSs may not simultaneously satisfy the inferred $\Lambda$ range while occupying the lower mass--gap region (see Appendix~\ref{app:noDM_CFL}).  {It should be noted that the $\Lambda$ limits derived from GW170817 were originally inferred within single--fluid compact star frameworks; therefore, their application to the present two--fluid configurations provides only qualitative guidance rather than definitive quantitative constraints.}} Additionally, we study the spatial distribution of dark matter within the star, which may appear  as either a halo-like or core-like configuration. We further investigate how decreasing the dark matter radius relative to the quark-matter radius influences $\Lambda_{1.4\,M_\odot}$. We acknowledge that the topics such as SQSs and dark matter have been subjects of extensive researches. However, the present manuscript offers several innovative aspects that distinguish it from previous works:
	While other references such as Ref. \cite{Panotopoulos2017b,LopesDas} have explored the inclusion of dark matter in SQS models, our work goes further by implementing advanced corrections using Perturbative Quantum Chromodynamics (PQCD). This includes considering the running of the coupling constant and the running strange quark mass, based on the latest Particle Data Group  data-set. This level of theoretical precision in modeling SQS with dark matter is less commonly observed in prior studies. Moreover, a significant innovative contribution is demonstrating that for each fixed gap parameter, there exists a critical dark matter mass beyond which further increases in this critical value lead to a reduction in the maximum observable mass ($M_{TOV}$). Identifying this critical point provides deeper insight into how dark matter influences the structure of SQS. {Notably, we demonstrate that a stiff EOS, which might seem incompatible with $\Lambda$ bounds in a single-fluid model, can be brought into {qualitative alignment} with these observations using a two-fluid framework. This finding is the main motivation of our study.}
	
	The structure of the paper is organized as follows. In Section~\ref{mass and coupling}, we incorporate perturbative QCD (PQCD) corrections by including the running coupling constant and running strange quark mass, based on the latest Particle Data Group dataset. In Section~\ref{cflphase}, we provide an overview of the CFL phase of quark matter and discuss its implications for the EOS through the framework of color superconductivity. Section~\ref{e-p} is devoted to the construction of the thermodynamic potential for SQM.  Section~\ref{BECcond} introduces the dark matter model, where it is treated as a Bose--Einstein condensation, and presents the corresponding EOS. In Section~\ref{results}, we first outline the two--fluid formalism used to solve the
	Tolman--Oppenheimer--Volkoff equations and derive the $\Lambda$ parameter.
	We then present  our numerical results for mass--radius ($M$--$R$)
	relations, $\Lambda$ versus mass ($\Lambda$--$M$), and the spatial distribution of dark
	matter, highlighting the impact of the halo radius on $\Lambda$.
	Finally, we revisit the conventional interpretation of the mass--gap in compact objects and discuss the implications of our results. 
	Section~\ref{critmD}  discusses the existence of a critical dark matter mass \( m_D \) that separates two distinct structural regimes in SQS configurations. At the end of  paper, we summarize the main findings and discuss the concluding remarks of the paper.
	
	\section{Running coupling and running strange quark mass based on the latest Particle Data Group dataset} \label{mass and coupling}
\begin{figure}[h!]
		\centering
		\par
		\includegraphics[width=7cm]{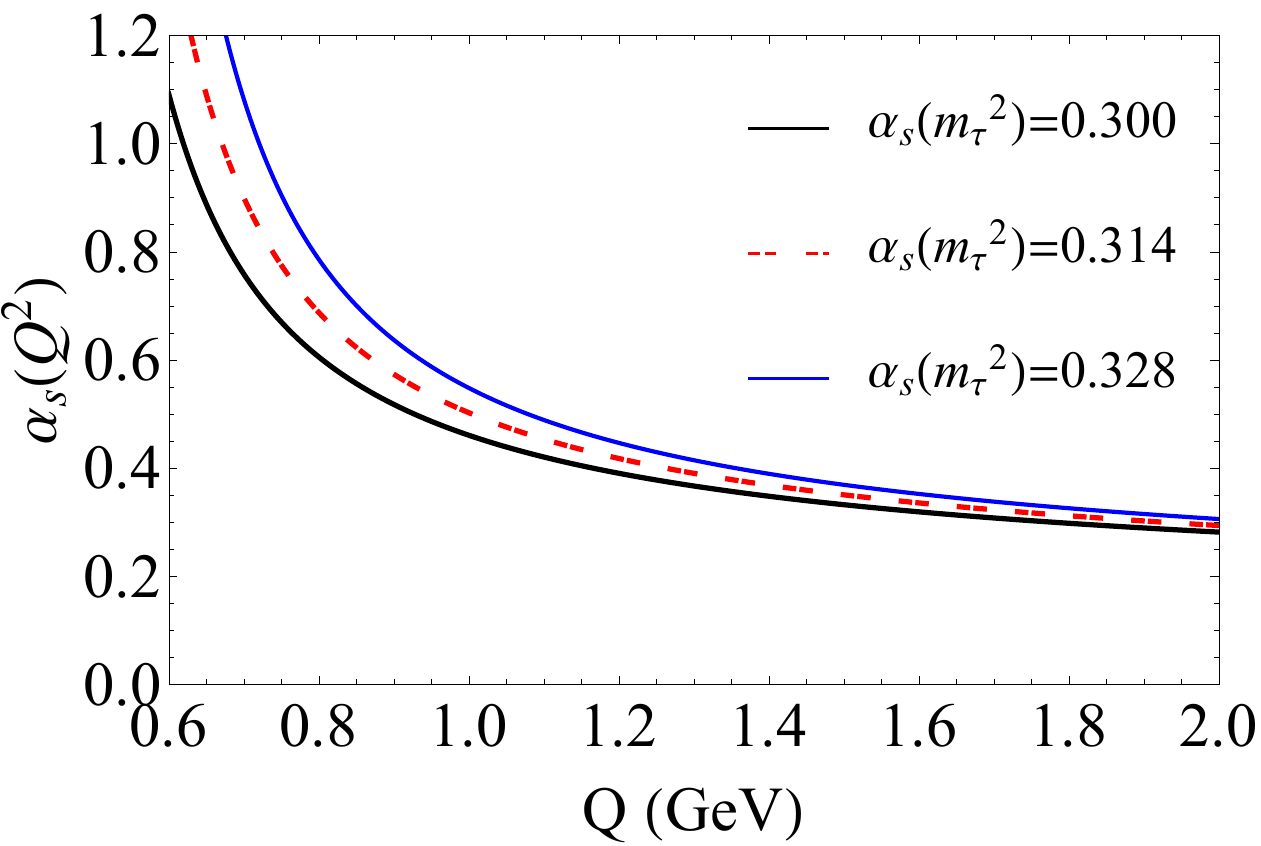}
		\caption{Running coupling constant of QCD as a function of energy for various values of $\alpha_s(m_\tau^2)$.}
		\label{couplingdiagram}
	\end{figure}

\begin{figure}[h!]
		\centering
		\par
		\includegraphics[width=8.5cm]{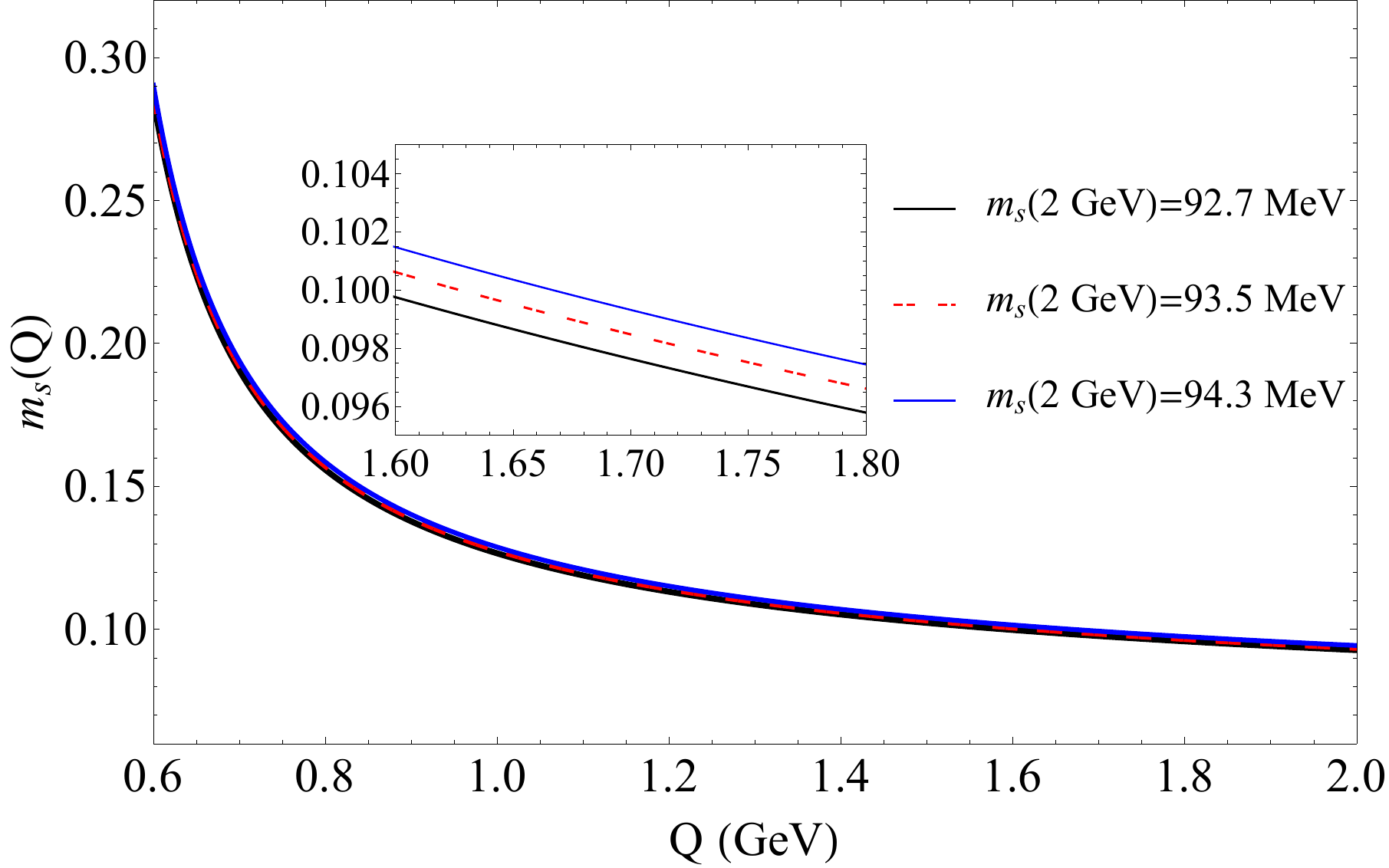}
		\caption{Running mass of the strange quark as a function of energy for different values of $m_s$(2GeV).}
		\label{running mass}
	\end{figure}

	As it is mentioned in the previous section, we use PQCD to derive the EOS of the SQM. In contrast to most previous studies on two--fluid models composed of SQM and dark matter--where the MIT bag model is commonly employed for describing the quark matter sector \cite{Panotopoulos2017b,LopesDas}--employing PQCD  offers distinct theoretical advantages. Firstly, the PQCD model is rooted in the renormalizable structure of the Standard Model, ensuring finite and physically meaningful predictions. Secondly, unlike the bag model where the speed of sound remains fixed, the perturbative approach naturally accounts for its energy dependence, allowing a more realistic treatment of the thermodynamic behavior at high densities \cite{sedaghatannals}. Additionally, in the PQCD framework, both the running coupling constant and the strange quark mass evolve with energy, in contrast to the bag model where these parameters are treated as constants. This energy dependence plays a crucial role in accurately modeling the EOS for SQM, particularly under conditions relevant to compact stars. In perturbative analyses, it is standard to express physical observables as a series expansion in the coupling parameter. Consequently, our preliminary analysis investigates the behavior of the QCD coupling constant, $\alpha_s$, as a function of energy. The dependence of $\alpha_s$ on the renormalization scale $Q$ is governed by the following expression
	\cite{Vermaseren,Fraga2006}, 
	\begin{equation} \label{QCD coupling}
		\alpha_s (Q) =\frac{4 \pi  \left(1-\frac{2\beta_1 \log (L)}{{\beta_0}^2 L}\right)}{\beta_0L},
	\end{equation}
	where $\beta_0= 11 - 2\frac{N_f}{3}$, $\beta_1=51 - 19\frac{N_f}{3}$,  $L=2\log(\frac{Q}{\Lambda_{\overline{MS}}})$, and $N_f$ denotes the number of flavors, which is taken to be $3$ in our calculations. The parameter $\Lambda_{\overline{MS}}$ corresponds to the renormalization scale within the modified minimal subtraction scheme and is determined using data from the Particle Data Group 2024 dataset \cite{Navas2025} by ensuring that $\alpha_s(m_{\tau})=0.314^{+0.014}_{-0.014}$. $m_{\tau}$ is the mass of the tau lepton, which is adopted to $1776.86 MeV$ \cite{Navas2025}. The evolution of $\alpha_s$ as a function of $Q$ for various initial values of $\alpha_s(m_\tau^2)$ is illustrated in Fig. \ref{couplingdiagram}. 
	As shown in Fig. \ref{couplingdiagram}, the variation in $\alpha_s$ becomes narrower with increasing energy.  We set $\alpha_s(m_\tau) = 0.314$ for the rest of our analysis.
	In this study, we consider baryonic matter to consist of SQM, which includes up, down, and strange quarks, evaluated at zero temperature and nonzero chemical potential. Within this framework, the masses of the up and down quarks are considered negligible in comparison to that of the strange quark. To leading order, the variation of the strange quark mass with energy scale, denoted by $m_s(Q)$, is governed by the following expression
	\cite{Vermaseren,Fraga2006},
	\begin{equation}
		m_{s}(Q)=m_{s}(2GeV)\left[ \dfrac{\alpha _{s}(Q)}{\alpha _{s}(2GeV)}\right]
		^{\dfrac{\gamma _{0}}{\beta _{0}}},  \label{9}
	\end{equation}%
	where, $\gamma_0$ represents the anomalous dimension and is given by $\gamma_0 = 3\tfrac{N_c^2 - 1}{2N_c}$, where $N_c = 3$ denotes the number of color charges. Based on the most recent data from the Particle Data Group, the strange quark mass at a scale of 2GeV, noted as $m_s(2\text{GeV})$, is reported to be $93.5^{+0.8}_{-0.8}$ MeV. For our analysis, we adopt the central value, $m_s(2\text{GeV}) = 93.5$ MeV.
	The energy dependence of $m_s(Q)$ for several input values of $m_s(2\text{GeV})$ is depicted in Fig. \ref{running mass}.  
	
	As previously stated, we explore SQS in the CFL phase. The following section provides a comprehensive overview of the CFL phase of quark matter and  its effects on the EOS within the context of color superconductivity.

	\section{color--flavor--locked phase}\label{cflphase}
	
	At extremely high densities, quark matter is predicted to exhibit color superconductivity, where quarks form Cooper pairs in a manner similar to electron pairing in conventional superconductors \cite{Alford2008}. This phenomenon stems from the strong interaction, which remains attractive in specific color channels, causing spontaneous color symmetry breaking and the emergence of a superconducting phase \cite{Alford2008,Rajagopal2001}. In environments of ultra-dense matter, the CFL phase is the most favored state, where up, down, and strange quarks pair symmetrically, reducing the original SU(3) color and flavor symmetries to a common diagonal subgroup \cite{Shovkovy2005,Schmitt2006,Gholami2025}. The presence of color superconductivity profoundly changes the EOS of quark matter, influencing the mass-radius relation of compact stars and their transport properties \cite{Kurkela2010,Sedaghat2021, Alford2003}. Although color superconductivity is strongly supported by QCD theory, it has not been directly observed, because the extreme densities required for its formation cannot be recreated in laboratory experiments. However, indirect clues from astrophysical observations -- such as how neutron stars cool, their rotational behavior, and possible neutrino signals -- offer potential ways to detect this phase \cite{Shovkovy2005, Schmitt2006, Page2009}.
	
	In the next section, we derive the thermodynamic potential of SQM in the CFL phase to evaluate its thermodynamic properties.

	\section{Thermodynamic potential of strange quark matter}\label{e-p}
	As stated in the introduction, the goal of this paper is to investigate a compact star composed of two--fluid matter: SQM and scalar dark matter. In this section, we outline the method for calculating the thermodynamic potential of SQM, $\Omega_{\text{SQM}}$,  in CFL phase.
	To formulate $\Omega_{\text{SQM}}$, we separate it into three distinct contributions. The first corresponds to the free (non-interacting) component, which includes non-interacting quarks and electrons. The second accounts for interactions arising from QCD at leading order, and the third term is the term accounting
	for the condensation energy of Cooper pairs in the CFL phase. The QCD interaction term, originally obtained through a two-loop Feynman diagram analysis, can be found in Refs. \cite{Fraga2006, Kurkela2010}. Below, we present the leading-order expression for the thermodynamic potential, $\Omega_{\text{LO}}$, of a system containing quark flavors with masses $m_f$ and chemical potentials $\mu_f$ for $f = u, d, s$, in addition to electrons characterized by the chemical potential $\mu_e$.
	\begin{equation}
		-\frac{\Omega_{\text{LO}}}{V}=\sum_{{N_f=1}}^{3}\left({\mathcal M}_1+\frac{{\mathcal M}_2\alpha_s(Q)}{4\pi}\right),
	\end{equation}
	where, $\mathcal{M}_1$ represents the non-interacting contribution, while the term $\frac{\mathcal{M}_2 \, \alpha_s(Q)}{4\pi}$ accounts for the PQCD corrections. The explicit forms of $\mathcal{M}_1$ and $\mathcal{M}_2$ are given below,
	\begin{equation}
		{\mathcal M}_1=
		\frac{N_c {\mu_f}^4}{24\pi^2}\bigg\{2\hat{u_f}^3-3z_f \hat{m_f}^2\bigg\} +\frac{{\mu_e}^4}{12\pi^2},\label{m1}
	\end{equation}
	\begin{equation}
		{\mathcal M}_2= \frac{d_A {\mu_f}^4}{4\pi^2}\Bigg\{-6z_f \hat{m_f}^2 \ln\frac{Q}{m} +2\hat{u_f}^4 - 4z_f \hat{m_f}^2 -3{z_f}^2 \Bigg\}, \label{m2}
	\end{equation}
	where $\hat{u_f}\equiv(\sqrt{{\mu_f}^2-m_f^2})/\mu_f$, $\hat{m_f}\equiv m_f/\mu_f$, $z_f \equiv \hat{u_f}-\hat{m_f}^2\,\ln\bigg[\frac{1+\hat{u_f}}{\hat{m_f}}\bigg]$ and $d_A\equiv {N_c}^2-1$. Therefore, the $\Omega_{\text{SQM}}$ is given by,
	\begin{equation}
		-\frac{\Omega_{\text{SQM}}}{V}=-\frac{\Omega_{\text{LO}}}{V}-\frac{\Omega_{\text{CFL}}}{V},
	\end{equation}
	where $-\frac{\Omega_{\text{CFL}}}{V}$ represents the contribution from the CFL phase, which is given by $\frac{\Delta^2 (\mu_u + \mu_d + \mu_s)^2}{3\pi^2}$, where $\Delta$ is the gap parameter \cite{Kurkela2010}.
	The condition of beta equilibrium imposes the following relationships among the chemical potentials,
	\begin{eqnarray}
		\mu _{s}&=&\mu _{d}\equiv \mu ,\nonumber\\\mu _{u}&=&\mu -\mu _{e}.  \label{24}
	\end{eqnarray}
	In accordance with the analysis of stable SQM provided in Ref. \cite{Kurkela2010}, the quantity $Q$ is taken to be $\frac{4}{3}(\mu_u + \mu_d + \mu_s)$ throughout this work. The pressure can be obtained through the expression $P = -B - \frac{\Omega_{\text{SQM}}}{V}$, where $B$ represents the contribution from non-perturbative dynamics not captured by perturbative calculations \cite{Kurkela2010,Sedaghat2021,JSedaghat}. To satisfy the boundary condition that the total pressure vanishes at the stellar surface, appropriate values for $B$ are selected accordingly \cite{Kurkela2010,Sedaghat2021,JSedaghat,sedaghat}. In the absence of perturbative corrections, $B$ corresponds to the traditional bag constant. However, once perturbative effects are included, $B$ deviates from the standard MIT bag model values and is thus interpreted as an effective bag parameter.
	While the parameter $B$ is typically treated as a constant throughout the SQS, this assumption does not align with the asymptotic freedom behavior predicted by QCD. As demonstrated in Ref. \cite{sedaghatannals}, introducing a Gaussian-type density dependence for $B$ within the framework of PQCD yields more realistic results. This approach not only improves the behavior of the sound speed, but also it increases the maximum gravitational mass, all without conflicting with the $\Lambda$ constraint from GW170817. Therefore, in the present work, we adopt a density-dependent form of $B$ given by,
	\begin{equation}
		B=B_0e^{-a(\frac{n_B}{n_0}-1)^2}. \label{DDbag}
	\end{equation}
	In this expression, $n_B$ denotes the baryon number density, $B_0$ represents the bag pressure at the stellar surface, and $n_0$ corresponds to $n_B$ at that point. The parameter $a$ (with $a = 0.5$ in this work) controls the rate at which $B$  decreases from the surface toward the core of the star \cite{sedaghatannals}. To obtain realistic results, we assume $n_B > 2n_{\text{sat}}$, where $n_{\text{sat}}$ is the saturation density, which equals $0.16 \, \text{fm}^{-3}$. With this choice, $B_0$ will be approximately $90\frac{\text{MeV}}{\text{fm}^3}$ for our EOSs.
	By knowing the pressure, additional thermodynamic quantities, including the energy density and quark number densities, can be also determined. The energy density, denoted by $\epsilon$, is expressed as,
	\begin{equation}\label{EOS}
		\epsilon=\sum_{{N_f}}\mu_fn_f+\mu_en_e-P.
	\end{equation}
	The quark number density for a given flavor $f$, denoted by $n_f$, is defined as $n_f = \frac{\partial P}{\partial \mu_f}$, and the electron number density $n_e$ is calculated using $n_e = \frac{\mu_e^3}{3\pi^2}$. The relationship expressed in Eq.~(\ref{EOS}) leads to the pressure and energy density relation which is known as the EOS. It is well known that characterizing the EOS is vital for investigating thermodynamic aspects such as the sound speed and for modeling the structural features of SQS.

	{Finally it is worth mentioning that for computed EOS, the thermodynamic consistency condition,
\begin{equation}
P-n_B^2\frac{d(\epsilon/n_B)}{dn_B}=0,
\end{equation} 
should be checked. In this regards, we have found that this condition is satisfied with high accuracy in our calculations for all EOS's. 
Here, for a general discussion regarding the thermodynamic consistency in density-dependent quark matter models, we refer the reader to Refs.~\cite{Xu2015,Xia2014}.}
	
	\section{Bose--Einstein condensation of dark matter}\label{BECcond}
	
	When the thermal de~Broglie wavelength exceeds the mean inter-particle distance in a dilute Bose gas, the system undergoes a quantum phase transition into a Bose--Einstein condensation (BEC), where particles collectively occupy the ground state. Assuming the zero-temperature limit ($T = 0~\mathrm{K}$), the majority of dark matter particles reside in this condensed phase. In this regime, the dynamics are dominated by low-energy interactions, which can be effectively described by s-wave scattering with scattering length $l_a$. As a result, the detailed form of the interaction potential can be approximated by a mean repulsive interaction, significantly simplifying the treatment of the system \cite{Lopes2018,LopesDas,sedaghatonefluid},
	\begin{equation}
		V(\textbf{r}
		-\textbf{r}')=\frac{4\pi l_a}{m_{\text{D}}}\delta(\textbf{r}
		-\textbf{r}')
	\end{equation}
	The EOS corresponding to this condensed phase has been derived in Refs.~\cite{Panotopoulos2017b,Lopes2018}, which is given by,
	\begin{equation}\label{omegaBEC}
		-\frac{\Omega_{BEC}}{V}=P_{BEC}=\frac{2\pi l_a}{m_{\text{D}}^3}\epsilon_{BEC}^2,
	\end{equation}
	where, $P_{BEC}$ and $\epsilon_{BEC}$ refer to pressure and energy density of dark scalars in Bose--Einstein condensation. The s-wave scattering length $l_a$ is taken to be $1~\mathrm{fm}$
	\cite{Panotopoulos2017b,Lopes2018,LopesDas}. 
	\section{Results for structural properties}\label{results}
	In this section we explore the structural properties of SQS in the CFL phase in the presence of scalar dark matter.
	we first discuss the formulation and application of the Tolman-Oppenheimer-Volkoff (TOV) equations and $\Lambda$ in the context of a two--fluid model for compact stars. 
	\subsection{TOV equations in two--fluid model}
	To determine the macroscopic structure of compact stars, we solve the Tolman-Oppenheimer-Volkoff (TOV) equations  \cite{Tolman1939,Oppenheimer1939}. In the conventional one--fluid approach, it is assumed that all constituents of the star are in perfect equilibrium and can be described by a EOS. However, in the two--fluid model, the star is treated as having two separate parts, each with its own EOS, and they only affect each other through gravity. The two--fluid TOV equations involve coupled differential equations for each fluid's pressure and energy density, leading to a more complex description of compact stars \cite{Panotopoulos2017,Karkevandi2022}. The two--fluid formalism is as follows,
	\begin{eqnarray}\label{tovS}
		\frac{dP_{\text{S}}}{dr}&=& -\left( P_{\text{S}} +\epsilon_{\text{S}} \right) \frac{m(r)+4\pi r^{3}P}{r(r-2m(r))}\,\label{tovD1},\\ 
		\frac{dP_{\text{D}}}{dr}&=& -\left( P_{\text{D}} +\epsilon_{\text{D}} \right) \frac{m(r)+4\pi r^{3}P}{r(r-2m(r))}, \label{tovD}
	\end{eqnarray}
	where $P_S$ and $P_D$ represent the pressures associated with SQM and dark matter, respectively. The total pressure and total mass of the system are given by $P=P_S+P_D$ and $m(r)=m(r)_S+m(r)_D$, where $m(r)_S$ denotes the mass contribution from SQM and {$m(r)_D$}  corresponds to the dark matter component. To solve the TOV equations numerically, we must specify the central pressures for both SQM and dark matter, $P_S(0)$ and $P_D(0)$. {In the subsequent analysis, we present the $M$--$R$ relations for different values of the central dark matter pressure fraction, defined as
		\begin{equation}
		f_r = \frac{P_D(0)}{P_S(0) + P_D(0)} \, ,
		\end{equation}
		where $P_S(0)$ and $P_D(0)$ denote the central pressures of the SQM and dark matter components, respectively. 
		This parameter quantifies the relative contribution of dark matter to the total central pressure at the stellar center in the two-fluid configuration.}

	\subsection{Dimensionless tidal deformability}
	The detection of gravitational waves from binary neutron star mergers has made the $\Lambda$ parameter  an essential observable for studying compact stars. This dimensionless parameter reflects how much a star deforms under tidal forces from a companion, offering valuable information about its internal structure. It is especially useful for examining exotic matter, such as that possibly found in SQS. A larger $\Lambda$ means the star is more easily deformed, while a smaller $\Lambda$ indicates a more compact and rigid object \cite{LopesDas}. To determine $\Lambda$, we must solve  the metric function $H(r)$ using the following equation \cite{Hinderer, ChengMingLi2020, sedaghatonefluid,Karkevandi2022},
	\begin{eqnarray}
		\frac{d\beta}{dr}&=& 2(1-2\frac{m(r)}{r})^{-1}H\{-\frac{6\pi}{r^2}[5\epsilon+9P+\sum_{{i=S,D}}f_i(\epsilon_i+P_i)]\nonumber\\%
		&+&2(1-2\frac{m(r)}{r})^{-1}(\frac{m(r)}{r^2}+4\pi r P)^2\}\nonumber\\
		&+&\frac{2\beta}{r}(1-2\frac{m(r)}{r})^{-1} \nonumber\\
			&\times&\{\frac{m(r)}{r}+2\pi r^2(\epsilon-P)-1\}   \label{HbetaEq}.
		\end{eqnarray}
		In this framework, the total pressure and energy density are given by $P = P_D + P_S$ and $\epsilon = \epsilon_D + \epsilon_S$, respectively. The parameter $\beta = \frac{dH}{dr}$ denotes the radial derivative of the metric function, and $f_S = \frac{d\epsilon_S}{dP_S} , f_D = \frac{d\epsilon_D}{dP_D}$ characterizes the rate at which the energy densities of both components change with respect to their pressures. The parameter $\Lambda$ can then be calculated using the following relation,
		\begin{equation}\label{DtidalD}
			\Lambda =\frac{2}{3}k_{2}(\frac{R_{out}}{M})^{5},
		\end{equation}
		where, $k_2$ denotes the dimensionless tidal Love number corresponding to the quadrupole mode ($l = 2$) \cite{Postnikov,Lopez2022,Lenzi2023}. It is worth noting that, in order to obtain $\Lambda$, $R_{out}$ denotes the outermost radius of the star, which may correspond to either the {SQM radius}  or the dark matter radius, depending on which component extends farther outward \cite{Karkevandi2022}.
		The formula for calculating $k_2$ is as follows,
		\begin{align}
			k_{2}& =\frac{16\sigma ^{5}}{5}(1-2\sigma )^{2}\left[ 1+\sigma (y-1)-\frac{y%
			}{2}\right]  \notag \\
			& \times \left\{ 12\sigma \left[ 1-\frac{y}{2}+\frac{\sigma (5y-8)}{2}\right]
			\right.  \notag \\
			& +4\sigma ^{3}\left[ 13-11y+\sigma \left( 3y-2\right) +2\sigma ^{2}\left(
			1+y\right) \right]  \notag \\
			& +\left. 6(1-2\sigma )^{2}\left[ 1+\sigma (y-1)-\frac{y}{2}\right] \ln
			\left( 1-2\sigma \right) \right\} ^{-1/2},  \label{tln}
		\end{align}
		where, $\sigma$ is defined as $M/R_{out}$, and $y$ is expressed by,
		\begin{equation}\label{conSQS}
			y=\frac{R_{out} \beta(R_{out})}{H(R_{out})} - \frac{4\pi R_{out}^3 \epsilon_0}{M}.
		\end{equation}
		Here, $\epsilon_0$ denotes the energy density evaluated at the outermost radius of the configuration \cite{ChengMingLi2020,sedaghatonefluid}. When the SQM component defines the outer boundary, $\epsilon_0$ is generally nonzero at the stellar surface. In contrast, if the dark matter component forms an extended halo and sets the outermost radius, the corresponding energy density vanishes at that boundary, yielding $\epsilon_0 = 0$.
		To determine the  $\Lambda$ as a function of the stellar mass, we solve Eqs. (\ref{tovD1}) and (\ref{tovD}), the metric perturbation Eq. (\ref{HbetaEq}), and the mass continuity equation $\frac{dM}{dr} = 4\pi r^2 \epsilon$ simultaneously. {We then evaluate our model results against the $\Lambda$ range inferred from the GW170817 event, {treating these observational bounds as qualitative guidelines for our two-fluid configurations.}} 
		
		In the following, we investigate how scalar dark matter influences the structural properties of SQSs in the CFL phase. Phenomenologically, the choice of $ \Delta $ in the range of 200 to 300 MeV is supported by the results in Refs. \cite{H.Gholami2025, Roupas2021}, as it aligns well with astrophysical constraints on mass, radius, and  $\Lambda$. Astrophysical constraints on the dense-matter EOS, as discussed in Ref. \cite{Kurkela2024}, place an upper bound on $ \Delta $, which is set at 216 MeV.
		Our analysis considers three different values for  $\Delta = 150, 175,$ and $200\,\text{MeV}$. We calculate the structural properties for two representative values of the {central dark matter pressure fraction, $f_r$} ($5\%$ and $10\%$). 
		For each value of $f_r$, the calculations are carried out for two distinct dark matter particle masses in order to clearly demonstrate the dependence of $\Lambda$ on both $f_r$ and $m_D$. 
		We begin with the case $f_r = 5\%$, followed by a detailed analysis of the $f_r = 10\%$ configuration. {In the following results, it is important to note that all numerical values reported for the dark-matter component are obtained by adopting the dark matter EOS in Eq.~(\ref{omegaBEC}) with a fixed scattering length $l_a = 1\,\mathrm{fm}$. Consequently, the quoted values of $m_D$ are conditional on this choice and should be interpreted as model-dependent.}

		\subsection{Results for $f_r = 5\%$ and Set~1 of $m_D$: $m_D = 225, 250, 275\,\text{MeV}$}
		
		\begin{figure*}[htbp]
			\centering
			\par
			\includegraphics[width=18cm]{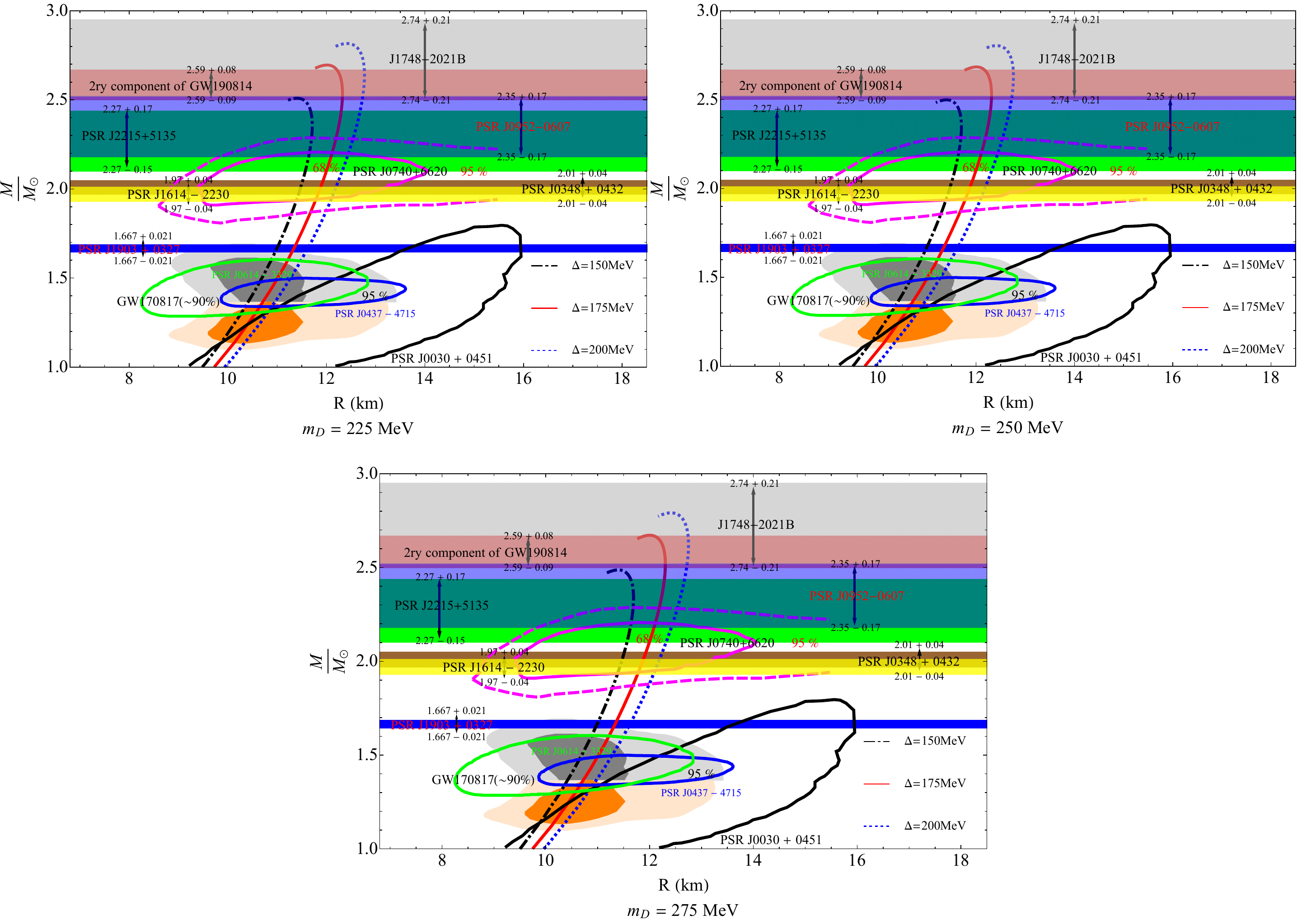}
\caption{ {$M$--$R$ relations for $f_r = 5\%$ and different values of $m_D$ and $\Delta$. 
	The shaded horizontal bands indicate the observational mass constraints from various compact objects, while the contour regions represent mass--radius constraints inferred from astrophysical observations. 
	The radius $R$ corresponds to the SQM component radius. 
	 {The GW170817 contour is included for qualitative comparison within the present two-fluid framework.}}}
			\label{mrdiagrams1}
		\end{figure*}
		We now present the results obtained for the first $m_D$ set for $fr=5\%$. This includes an analysis of the $M$--$R$ and $\Lambda$--$M$ relations, and the distribution of dark matter---whether it forms an extended halo or remains confined to the core. It should be emphasized that the radius shown in the \(M\!-\!R\) plots corresponds to the {SQM radius}, whereas \(R_{\text{out}}\) is used in the calculation of \(\Lambda\!-\!M\) plots. Since baryonic matter is responsible for electromagnetic interactions and the emission of observable signals such as thermal radiation and X-rays, the radius at which the pressure of the baryonic component vanishes is typically identified as the stellar radius \cite{Karkevandi2022,Zakary2024}. This definition is well aligned with observational measurements and enables direct comparison between theoretical models and observational data. 
		However, in the context of $\Lambda$, the relevant length scale is the outermost radius of the compact object. In two-fluid configurations, this outer radius may exceed the baryonic radius if the dark matter component forms an extended halo, and it is this outermost radius that enters the calculation of the $\Lambda$. Using Eqs.~ (\ref{tovS}) and (\ref{tovD}), we obtain the mass and corresponding stellar radius. We see that the $M$--$R$ relations derived from our two--fluid model---featuring SQS in the CFL phase admixed with scalar bosonic dark matter---demonstrate strong consistency with a wide range of astrophysical observations.  
		\begin{figure*}[htbp]
			\centering
			\par
			\includegraphics[width=18cm]{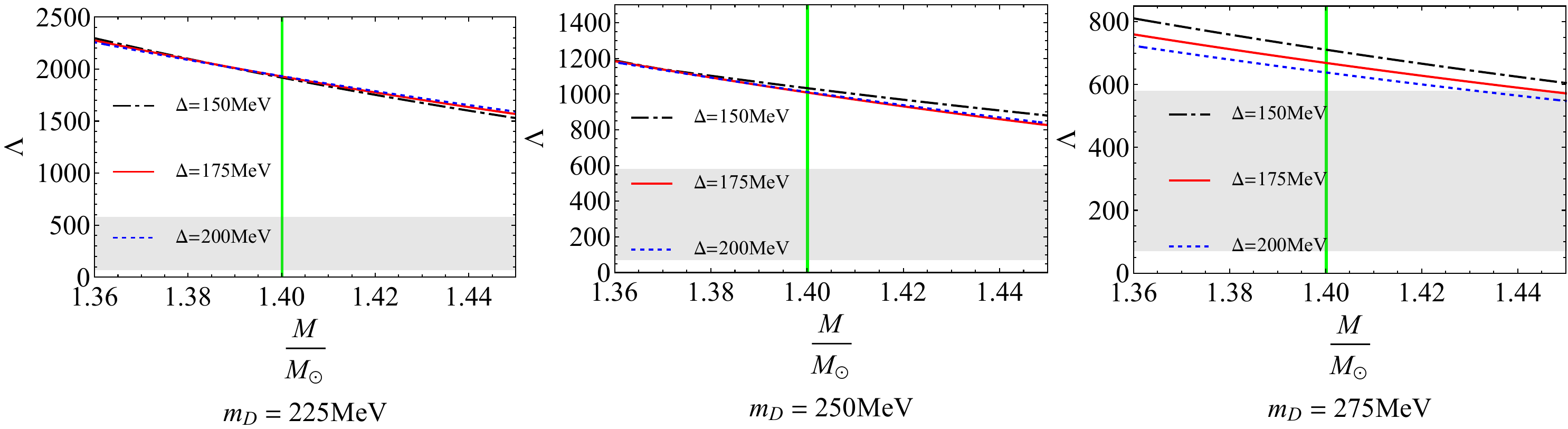}
			\caption{ {$\Lambda$ versus $M$ relations for $f_r = 5\%$ and different values of $m_D$ and $\Delta$. The shaded gray region represents the tidal-deformability range inferred from GW170817, utilized here as an  {indicative guideline} for our two-fluid model ($70 \lesssim \Lambda_{1.4M_\odot} \lesssim 580$).}}
			
			\label{tidaldiagrams1}
		\end{figure*}
		\begin{figure*}[htbp]
			\centering
			\begin{subfigure}{0.8\textwidth}
				\centering
				\includegraphics[width=\textwidth]{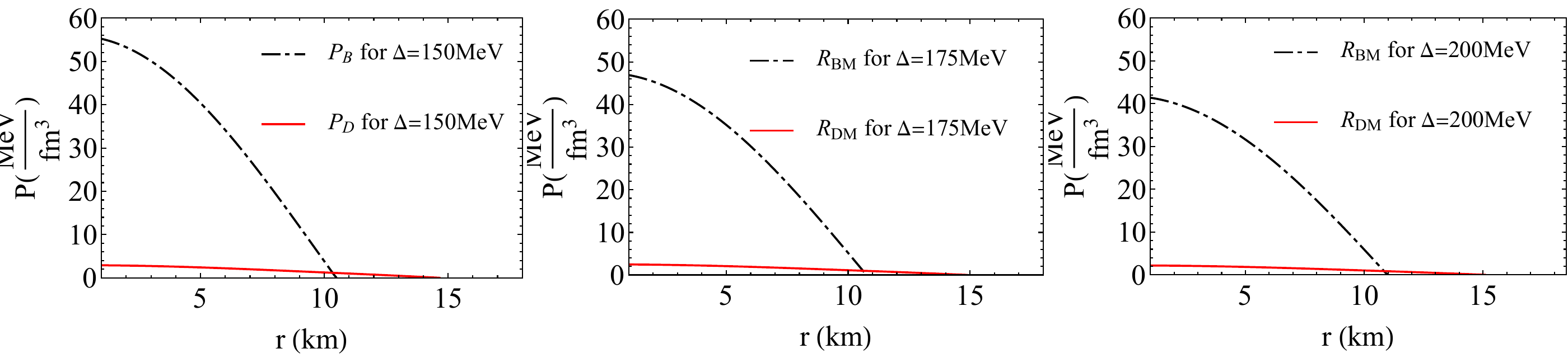}
				{$m_D=225MeV$}
			\end{subfigure}
			\hfill
			\begin{subfigure}{0.8\textwidth}
				\centering
				\includegraphics[width=\textwidth]{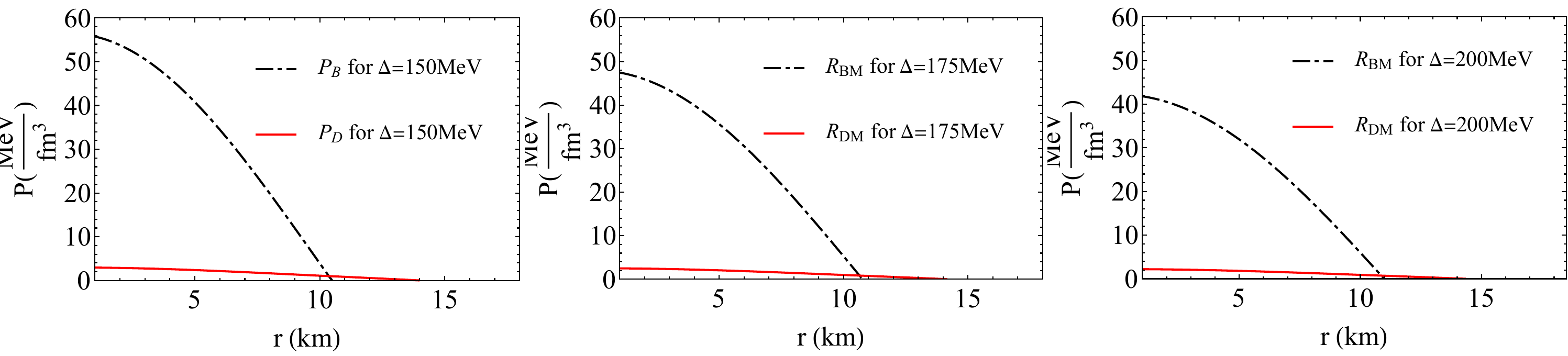}
				{$m_D=250MeV$}
			\end{subfigure}
			\hfill
			\begin{subfigure}{0.8\textwidth}
				\centering
				\includegraphics[width=\textwidth]{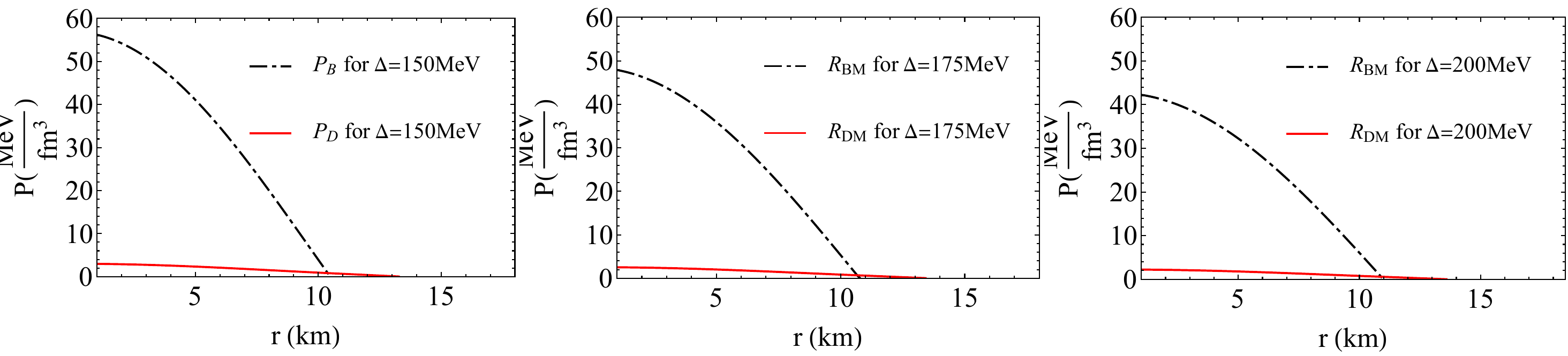}
				{$m_D=275MeV$}
			\end{subfigure}
			\caption{ $P_B(r)$ and  $P_D(r)$ corresponding to the  $M = 1.4 M_\odot$, for $fr = 5\%$. The plots include  different values of $m_D$ and  $\Delta$.}
			\label{pr1}
		\end{figure*}
		As shown in Fig. \ref{mrdiagrams1}, the $M-R$ diagrams cover several high-mass and ultra-massive compact objects. {Notably, our model reproduces the observed mass ranges of the following compact objects,}
		\begin{itemize}
			\item  {PSR J0952--0607}, an ultra-massive pulsar with a measured mass of $2.35 \pm 0.17\, M_\odot$ \cite{J0952}.
			\item {PSR J2215+5135}, a high-mass millisecond pulsar with an estimated mass around $2\, M_\odot$ \cite{Linares2018}.
			\item The secondary component of {GW190814}, with an inferred mass of $2.59\, M_\odot$, lying within the so-called mass--gap \cite{Abbott896}.
			\item The super-massive compact object   {PSR  J1748--2021B}, with a proposed mass of $2.74\, M_\odot$ \cite{Freire2008}.
			\item  {PSR J0740+6620}, constrained by NICER and XMM-Newton observations to have a mass around $2.1\, M_\odot$ and a well-bounded radius \cite{Cromartie2019}.
		\end{itemize}
	 {The $M\!-\!R$ diagrams also include the mass--radius contours inferred for PSR~J0030+0451~\cite{Riley2019}, 
		PSR~J0437--4715~\cite{PSR J0437 - 4715}, and PSR~J0614--3329~\cite{PSR J0614 - 3329}, together with the measured mass ranges of 
		PSR~J0348+0432~\cite{Antoniadis2013}, PSR~J1614--2230~\cite{Demorest2010}, and PSR~J1903+0327~\cite{Gangopadhyay}. 
		For reference, the GW170817 contour~\cite{Abbott2018} is included for qualitative comparison, 
		noting that the corresponding constraint was originally derived within single-fluid compact-star models. We now examine how our results relate to observationally inferred bounds on the $\Lambda$. 
		Based on the binary neutron star merger GW170817, the single-fluid parameter $\Lambda_{1.4M_\odot}$ is inferred to lie within 
		$70 \lesssim \Lambda_{1.4M_\odot} \lesssim 580$~\cite{Abbott2018}. 
		For completeness, we also note that if the secondary component of GW190814 is assumed to be a neutron star, the corresponding single-fluid estimate becomes 
		$458 < \Lambda_{1.4M_\odot} < 889$~\cite{Abbott896}. 
		However, in the present two-fluid framework we employ only the GW170817 interval--and only in a qualitative sense--as a reference guideline rather than a strict quantitative constraint.} Figure~\ref{tidaldiagrams1} shows that, for $f_r = 5\%$, the dark matter mass set $m_D = 225, 250, 275\,\mathrm{MeV}$ does not satisfy the observational constraint on $\Lambda$. However, the results indicate that increasing $m_D$ can lead to compliance with this constraint. Motivated by this behavior, we therefore consider a second set of $m_D$ with larger values. Before doing so,
		however, we first analyze the origin of the observed behavior of $\Lambda$ for the first
		set of dark matter masses. In particular, we investigate whether the dark matter component
		forms a compact core or instead extends outward to produce a halo-like structure SQM. 
		Figure~\ref{pr1} illustrates the baryonic matter radius ($R_{\mathrm{SQM}}$) and the dark
		matter radius ($R_{\mathrm{DM}}$) for a stellar configuration with mass
		$1.4\,M_\odot$. Specifically, this figure shows the radial profiles of the baryonic pressure
		($P_B$) and the dark matter pressure ($P_D$) as functions of the radial coordinate $r$.
		The radii $R_{\mathrm{SQM}}$ and $R_{\mathrm{DM}}$ are defined by the locations at which
		$P_B$ and $P_D$ vanish, respectively. Although such pressure profiles can be obtained for
		any choice of central pressure, the profiles displayed in Fig.~\ref{pr1} correspond to the
		central pressure that yields a total mass of $1.4\,M_\odot$.
		It is evident from the figure that, for all configurations considered,
		$R_{\mathrm{DM}}$ is significantly greater than $R_{\mathrm{SQM}}$, indicating the
		formation of an extended dark matter halo. Since the $\Lambda$ scales as
		$\Lambda \propto R_{\mathrm{out}}^{5}$, this large outer radius leads to {configurations favor a higher $\Lambda$ that is not qualitatively consistent with the current observational bounds.} 
		Furthermore, Table .~\ref{results1} and Fig.~\ref{pr1} indicate that for a fixed value of $\Delta$, increasing $m_D$ results in a moderate
		decrease in $M_{\mathrm{TOV}}$ accompanied by a significant reduction in the halo size
		or $R_{\mathrm{out}}$.
		These trends can be understood from the underlying microphysics. As shown in
		Eq.~(\ref{omegaBEC}), the dark matter pressure scales as $m_D^{-3}$. As a result,
		increasing $m_D$ suppresses the pressure contribution of the dark matter component,
		while gravitational attraction becomes more effective at larger $m_D$. This combination
		yields configurations that are more strongly gravitationally bound but less supported
		by pressure, leading to a lower value of $M_{\mathrm{TOV}}$. Moreover, heavier dark
		matter particles provide weaker pressure support, which enhances the central
		concentration of dark matter and explains the pronounced reduction in $R_{\mathrm{DM}}$.
		In light of these behaviors, we therefore consider a second set of
		dark matter masses, $m_D = 300, 325, 350\,\mathrm{MeV}$, and examine the corresponding
		results.
			
		\begin{table}[htbp]
			\caption{		
				Structural properties of SQS for $fr=5\%$ and different values of $m_D$ and  $\Delta$. 
{Here, $R$ and  $R_{1.4M_\odot}$ are the  {SQM radius} corresponding to the $M_{TOV}$ and $M=1.4M_\odot$, respectively.}}
			\centering
			\small
			\begin{subtable}{1\linewidth}
				\centering
				\scalebox{0.8}{
					\begin{tabular}{|c|c|c|c|c|c|c|}
						\hline
						\multicolumn{6}{|c|}{{$m_{\text{D}}$=225 MeV}} \\
						\hline
						$\Delta(MeV)$ & $\Lambda_{1.4\textup{M}_\odot}$& $R(km)$ & $M_{\text{TOV}}(\textup{M}_\odot)$ & $\frac{R_{\text{DM}}}{R_{\text{SQM}}}$ for $M=1.4\textup{M}_\odot$&$
{R_{1.4M_{\odot}}(km)}$ \\\hline
						$150$ & 1913.07 & 11.40 & 2.51 &  1.410 & {10.49}\\ \hline
						$175$ & 1932.67 & 12.02 & 2.70 &  1.389 & {10.80} \\ \hline
						$200$ & 1929.53 & 12.43 & 2.82 &  1.382 & {11.04}\\ \hline
				\end{tabular}}
				
			\end{subtable}
			\\
			\begin{subtable}{1\linewidth}
				\centering
				\scalebox{0.8}{
					\begin{tabular}{|c|c|c|c|c|c|}
						\hline
						\multicolumn{6}{|c|}{{$m_{\text{D}}$=250 MeV}} \\
						\hline
						$\Delta(MeV)$ & $\Lambda_{1.4\textup{M}_\odot}$& $R(km)$ & $M_{\text{TOV}}(\textup{M}_\odot)$ &  $\frac{R_{\text{DM}}}{R_{\text{SQM}}}$ for $M=1.4\textup{M}_\odot$&$
{R_{1.4M_{\odot}}(km)}$ \\\hline
						$150$ & 1032.41 & 11.40 & 2.50 &  1.343 & {10.49}\\ \hline
						$175$ & 1008.03 & 12.01 & 2.68 &  1.324 & {10.79}\\ \hline
						$200$ & 1003.02 & 12.42 & 2.80 &  1.310 & {11.05}\\ \hline
				\end{tabular}}
				
			\end{subtable}
			\\
			\begin{subtable}{1\linewidth}
				\centering
				\scalebox{0.8}{
					\begin{tabular}{|c|c|c|c|c|c|}
						\hline
						\multicolumn{6}{|c|}{{$m_{\text{D}}$=275 MeV}} \\
						\hline
						$\Delta(MeV)$ & $\Lambda_{1.4\textup{M}_\odot}$& $R(km)$ & $M_{\text{TOV}}(\textup{M}_\odot)$ &  $\frac{R_{\text{DM}}}{R_{\text{SQM}}}$ for $M=1.4\textup{M}_\odot$&${R_{1.4M_{\odot}}(km)}$\\\hline
						$150$ & 712.23 & 11.39 & 2.49 &   1.277 & {10.49}\\ \hline
						$175$ & 668.79 & 12.00 & 2.67 &  1.260 & {10.79}\\ \hline
						$200$ & 640.20 & 12.41 & 2.79 &  1.245 & {11.05}\\ \hline
				\end{tabular}}
				
			\end{subtable}
			\label{results1}
		\end{table}	
		
		\begin{figure*}[htbp]
			\centering
			\par
			\includegraphics[width=18cm]{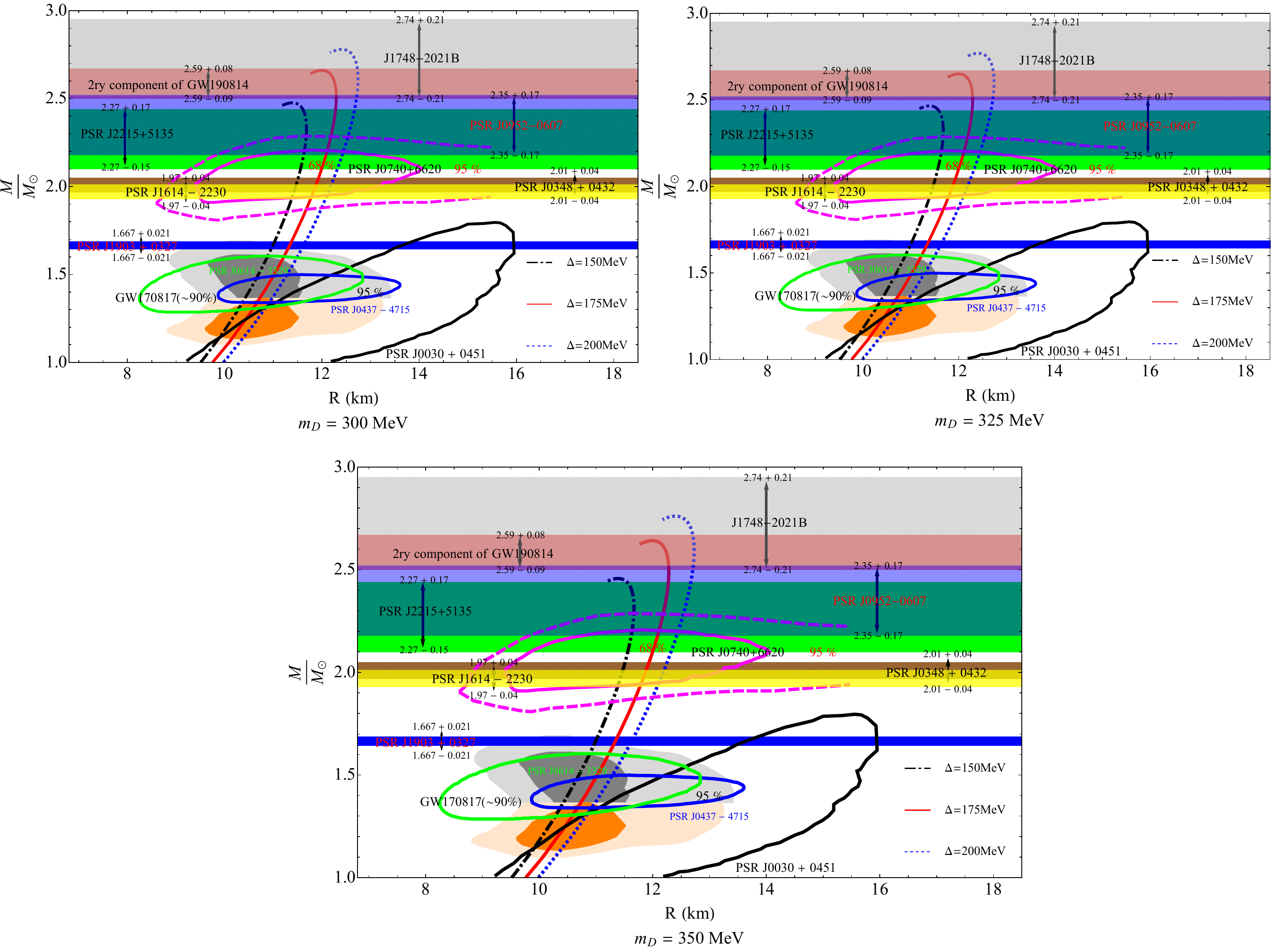}
		\caption{ {$M$--$R$ relations for $f_r = 5\%$ and different values of $m_D$ and $\Delta$. 
				The shaded horizontal bands indicate the observational mass constraints from various compact objects, while the contour regions represent mass--radius constraints inferred from astrophysical observations. 
				The radius $R$ corresponds to the SQM component radius. 
				{The GW170817 contour is included for qualitative comparison within the present two-fluid framework.}}}
			\label{mrdiagrams2}
		\end{figure*}
		
		\begin{figure*}[htbp]
			\centering
			\par
			\includegraphics[width=18cm]{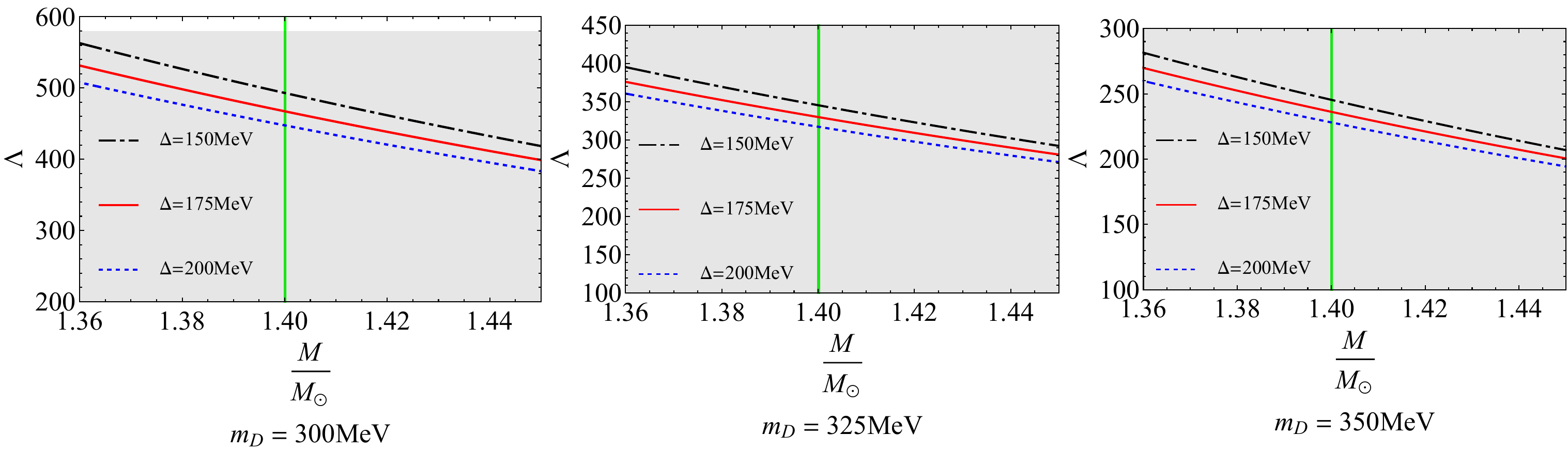}
					\caption{ {$\Lambda$ versus $M$ relations for $f_r = 5\%$ and different values of $m_D$ and $\Delta$. The shaded gray region represents the tidal-deformability range inferred from GW170817, utilized here as an  {indicative guideline} for our two-fluid model ($70 \lesssim \Lambda_{1.4M_\odot} \lesssim 580$).}}
			\label{tidaldiagrams2}
		\end{figure*}
		\begin{figure*}[htbp]
			\centering
			\begin{subfigure}{0.8\textwidth}
				\centering
				\includegraphics[width=\textwidth]{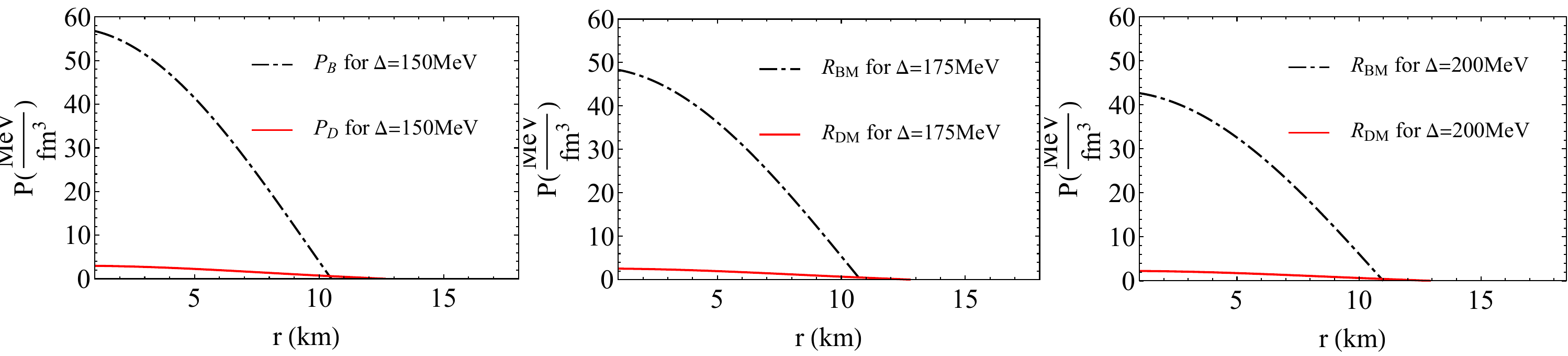}
				{$m_D=300MeV$}
			\end{subfigure}
			\hfill
			\begin{subfigure}{0.8\textwidth}
				\centering
				\includegraphics[width=\textwidth]{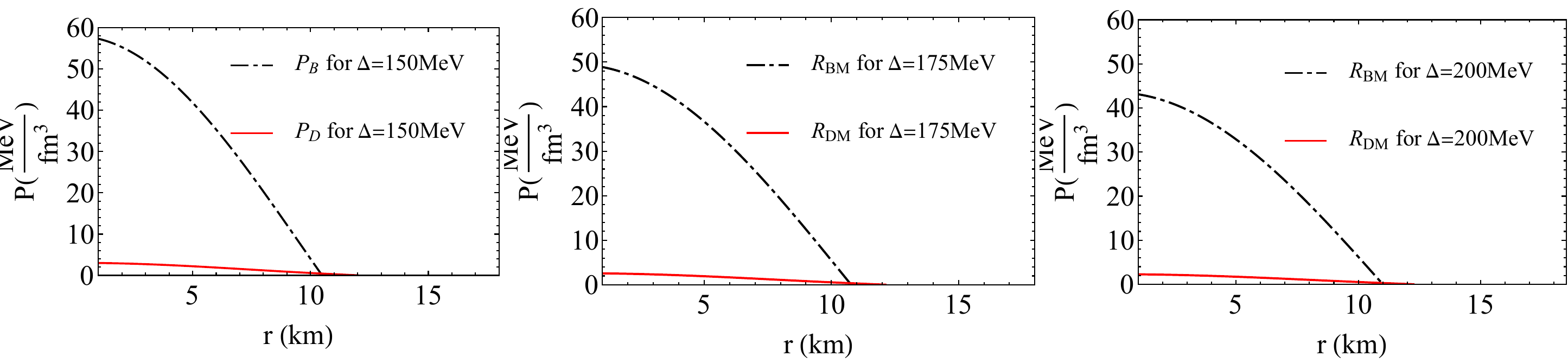}
				{$m_D=325MeV$}
			\end{subfigure}
			\hfill
			\begin{subfigure}{0.8\textwidth}
				\centering
				\includegraphics[width=\textwidth]{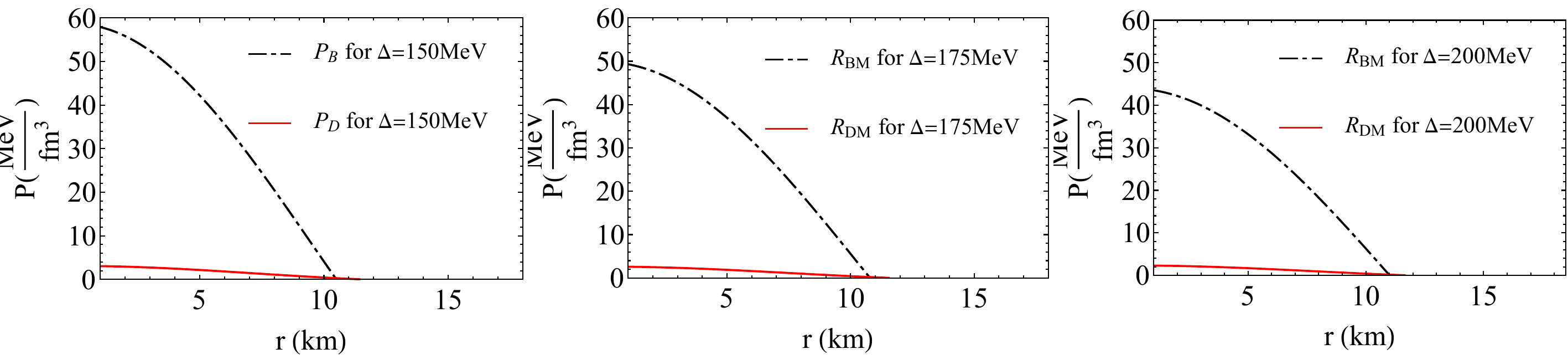}
				{$m_D=350MeV$}
			\end{subfigure}
			\caption{ $P_B(r)$ and  $P_D(r)$ corresponding to the  $M = 1.4 M_\odot$, for $fr = 5\%$. The plots include  different values of $m_D$ and  $\Delta$.}
			\label{pr2}
		\end{figure*}
		
		\begin{table}[htbp]
			\caption{		
				Structural properties of SQS for $fr=5\%$ and different values of $m_D$ and  $\Delta$. {Here, $R$ and  $R_{1.4M_\odot}$ are 
the {SQM radius} corresponding to the $M_{TOV}$ and $M=1.4M_\odot$, respectively.}}
			\centering
			\small
			\begin{subtable}{1.0\linewidth}
				\centering
				\scalebox{0.8}{
					\begin{tabular}{|c|c|c|c|c|c|}
						\hline
						\multicolumn{6}{|c|}{ {$m_{\text{D}}$=300 MeV}} \\
						\hline
						$\Delta(MeV)$ & $\Lambda_{1.4\textup{M}_\odot}$& $R(km)$ & $M_{\text{TOV}}(\textup{M}_\odot)$ & $\frac{R_{\text{DM}}}{R_{\text{SQM}}}$ for $M=1.4\textup{M}_\odot$&${R_{1.4M_{\odot}}(km)}$\\\hline
						$150$ & 492.72 & 11.39 & 2.48 &  1.219 & {10.49}\\ \hline
						$175$ & 467.54 & 12.00 & 2.66 &  1.194 & {10.80}\\ \hline
						$200$ & 447.44 & 12.41 & 2.78 &  1.182 & {11.06}\\ \hline
				\end{tabular}}
				
			\end{subtable}
			\\
			\begin{subtable}{1.0\linewidth}
				\centering
				\scalebox{0.8}{
					\begin{tabular}{|c|c|c|c|c|c|}
						\hline
						\multicolumn{6}{|c|}{ {$m_{\text{D}}$=325 MeV}} \\
						\hline
						$\Delta(MeV)$ & $\Lambda_{1.4\textup{M}_\odot}$& $R(km)$ & $M_{\text{TOV}}(\textup{M}_\odot)$ &  $\frac{R_{\text{DM}}}{R_{\text{SQM}}}$ for $M=1.4\textup{M}_\odot$&${R_{1.4M_{\odot}}(km)}$\\\hline
						$150$ & 345.31 & 11.39 & 2.47 &  1.162 & {10.49}\\ \hline
						$175$ & 331.14 & 12.00 & 2.65 &  1.139 & {10.80}\\ \hline
						$200$ & 317.00 & 12.40 & 2.77 &  1.117 & {11.06}\\ \hline
				\end{tabular}}
				
			\end{subtable}
			\\
			\begin{subtable}{1.0\linewidth}
				\centering
				\scalebox{0.8}{
					\begin{tabular}{|c|c|c|c|c|c|}
						\hline
						\multicolumn{6}{|c|}{ {$m_{\text{D}}$=350 MeV}} \\
						\hline
						$\Delta(MeV)$ & $\Lambda_{1.4\textup{M}_\odot}$& $R(km)$ & $M_{\text{TOV}}(\textup{M}_\odot)$ &  $\frac{R_{\text{DM}}}{R_{\text{SQM}}}$ for $M=1.4\textup{M}_\odot$&${R_{1.4M_{\odot}}(km)}$\\\hline
						$150$ & 245.44 & 11.39 & 2.46 &   1.105 & {10.48}\\ \hline
						$175$ & 235.75 & 11.99 & 2.64 &  1.083 & {10.80}\\ \hline
						$200$ & 227.76 & 12.40 & 2.76 &  1.063 & {11.06}\\ \hline
				\end{tabular}}
				
			\end{subtable}
			\label{results2}
		\end{table}
		
		\subsection{Results for $f_r = 5\%$ and Set~2 of $m_D$: $m_D = 300, 325, 350\,\text{MeV}$}
		In this subsection, we examine the structural properties of SQS for the second set of
		dark matter masses, $m_D = 300, 325, 350\,\mathrm{MeV}$. Figures~\ref{mrdiagrams2} and
		\ref{tidaldiagrams2} present the corresponding $M$--$R$ and $\Lambda$--$M$ diagrams,
		respectively. We find that these configurations not only cover the
		previously mentioned massive objects,  {but also remain  {qualitatively consistent} with the $\Lambda$ constraints inferred from GW170817.}
		These results suggest that SQSs admixed with dark matter could naturally explain compact objects traditionally considered too massive to be neutron stars. The presence of the dark matter component provides additional pressure support, enabling the existence of stellar configurations that exceed the maximum mass predicted by conventional nuclear EOSs.
		Crucially, our findings have important implications for the interpretation of the mass--gap between neutron stars and black holes, typically considered to span the $2.5$--$5\,M_{\odot}$ range \cite{ApJ499,ApJ725,ApJ757}.  {Our results suggest that exotic compact-star configurations may populate part of the conventionally defined lower mass-gap region.}
		Our two--fluid model shows that hybrid stars containing both quark matter and dark matter can reach masses matching some of the heaviest compact objects observed to date. These findings point toward a potentially crucial role for dark matter in determining the structure and stability of ultra-massive compact stars. Table \ref{results2} presents the structural properties corresponding to the second set of $m_D$ values with $fr=5\%$. As indicated in the table, for a fixed value of $\Delta$, increasing $m_D$ results in a
		moderate decrease in $M_{\mathrm{TOV}}$ while leading to a significant reduction in
		$\Lambda_{1.4\,M_\odot}$, similar to the trend observed for the first set of $m_D$.
		Conversely, for a fixed value of $m_D$, increasing $\Delta$
		produces a significant increase in $M_{\mathrm{TOV}}$ accompanied by a moderate decrease
		in $\Lambda_{1.4\,M_\odot}$. The reason that configurations with the second set of $m_D$ values show  { {qualitative agreement}} with
		$\Lambda$ constraint can be understood in terms of the halo radius. As shown in
		Fig.~\ref{pr2} and Table \ref{results2}, when compared with the configurations corresponding to the first set of
		$m_D$, the dark matter halo radius decreases significantly, and this behavior has a
		strong impact on $\Lambda$. More specifically, while the variation in the stellar mass
		is relatively small, the change in  $R_{\mathrm{out}}$ is substantial.
		As a result, the compactness $M/R_{\mathrm{out}}$ increases markedly, which in turn
		leads to a significant reduction in $\Lambda$. 

	\subsection{Results for $f_r = 10\%$ and Set~1 of $m_D$: $m_D = 300, 325, 350\,\text{MeV}$}
	In addition to the case with the contribution of dark matter with $fr=5\%$, we extend our analysis to a higher dark matter admixture of $f_r = 10\%$. In the previous subsection, we showed that configurations with
	$m_D = 300, 325, 350\,\mathrm{MeV}$ and $f_r = 5\%$ can account for the observed massive
	objects in the mass--gap region while remaining  {qualitatively consistent} with the constraint on
	$\Lambda$. We now demonstrate that increasing $f_r$ leads to a significant change in the
	behavior of $\Lambda$.
	\begin{figure*}
			\centering
			\par
			\includegraphics[width=18cm]{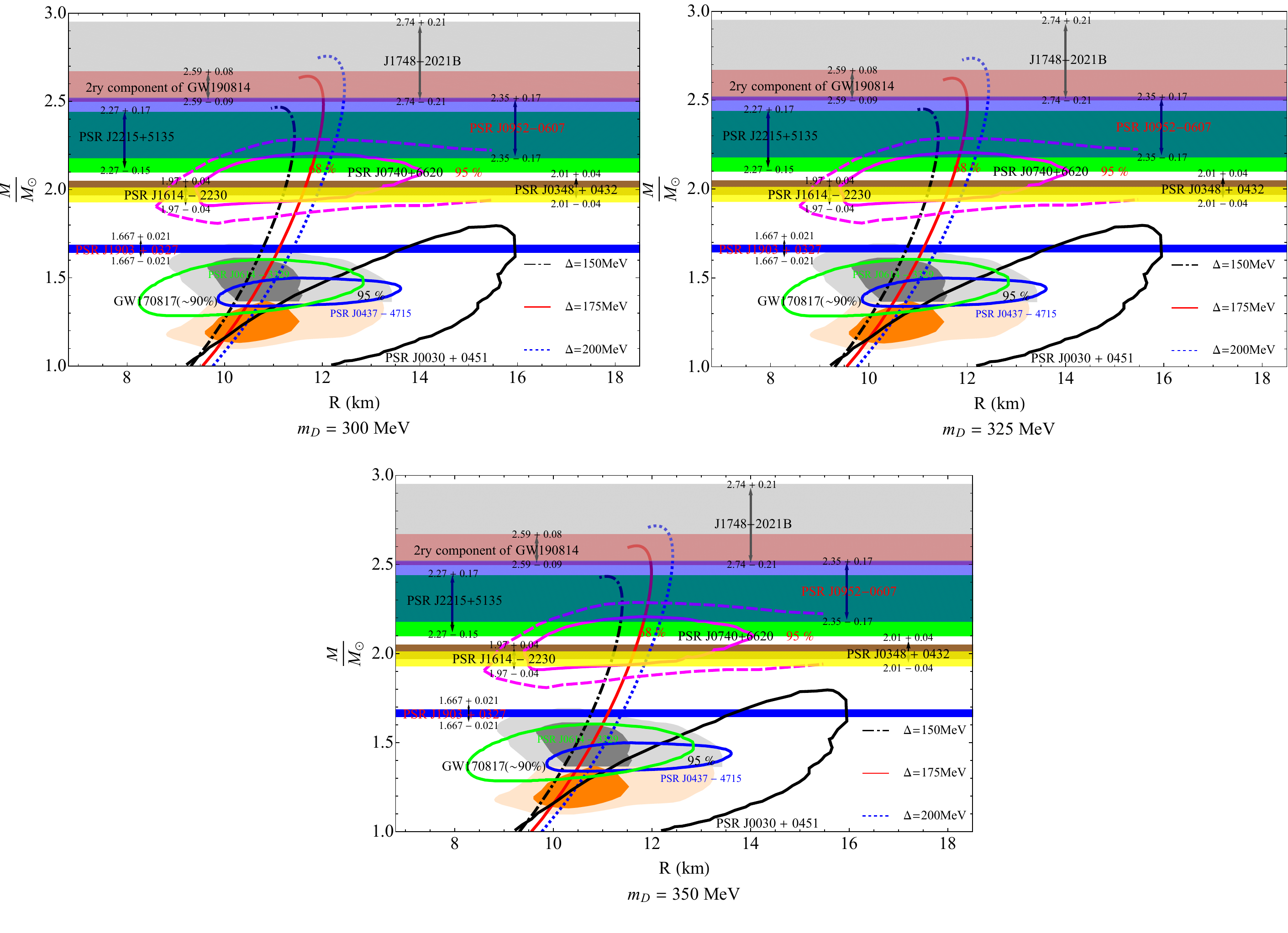}
			\caption{ {$M$--$R$ relations for $f_r = 10\%$ and different values of $m_D$ and $\Delta$. 
					The shaded horizontal bands indicate the observational mass constraints from various compact objects, while the contour regions represent mass--radius constraints inferred from astrophysical observations. 
					The radius $R$ corresponds to the SQM component radius. 
					 {The GW170817 contour is included for qualitative comparison within the present two-fluid framework.}}}
			\label{mrdiagrams3}
		\end{figure*}
		\begin{figure*}[htbp]
			\centering
			\par
			\includegraphics[width=18cm]{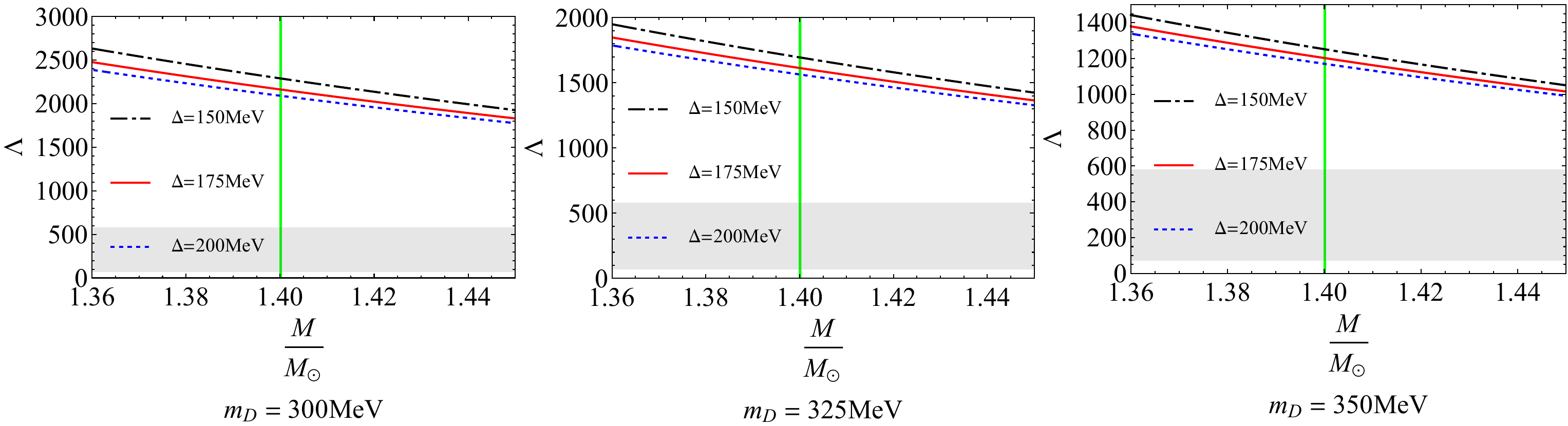}
					\caption{ {$\Lambda$ versus $M$ relations for $f_r = 10\%$ and different values of $m_D$ and $\Delta$. The shaded gray region represents the tidal-deformability range inferred from GW170817, utilized here as an  {indicative guideline} for our two-fluid model ($70 \lesssim \Lambda_{1.4M_\odot} \lesssim 580$).}}
			\label{tidaldiagrams3}
		\end{figure*}
		\begin{figure*}[htbp]
			\centering
			\begin{subfigure}{0.8\textwidth}
				\centering
				\includegraphics[width=\textwidth]{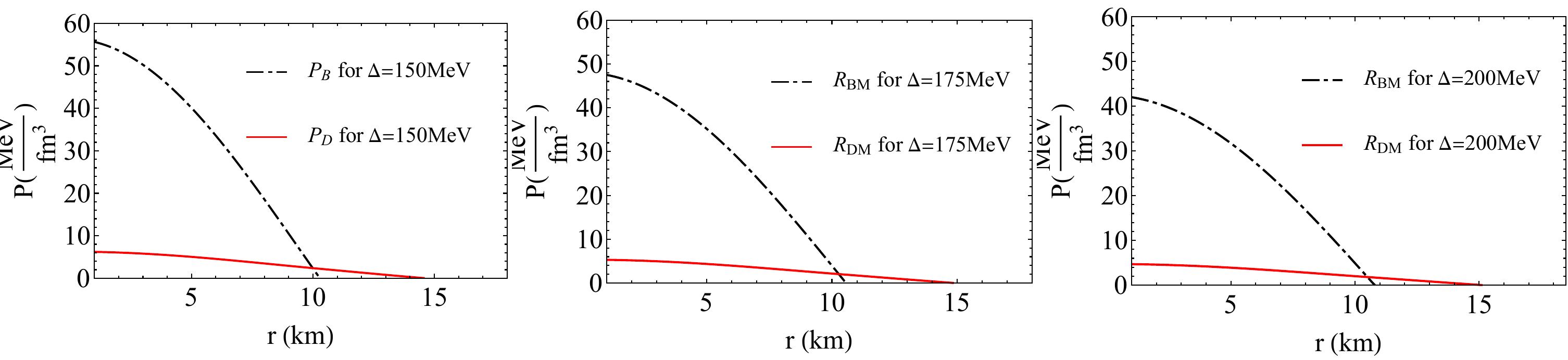}
				{$m_D=300MeV$}
			\end{subfigure}
			\hfill
			\begin{subfigure}{0.8\textwidth}
				\centering
				\includegraphics[width=\textwidth]{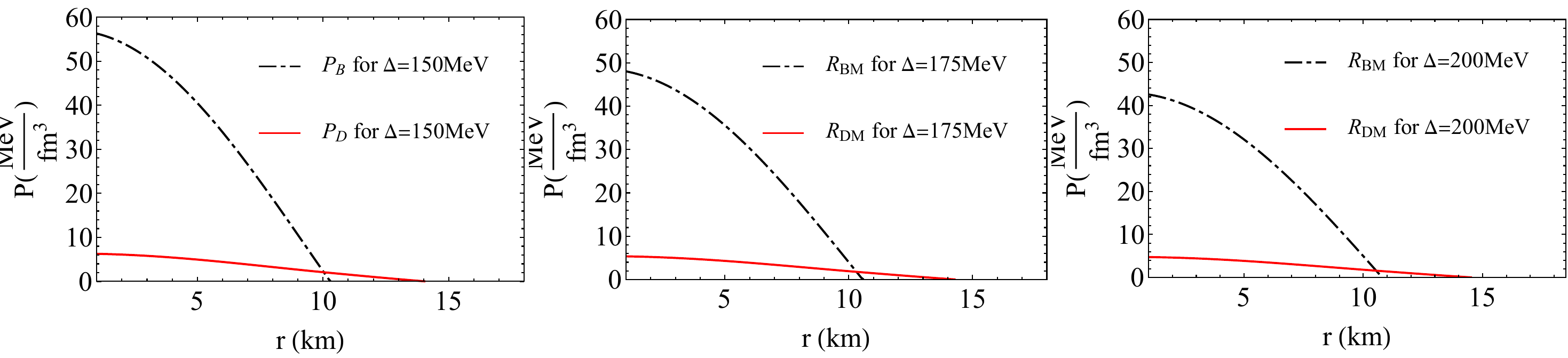}
				{$m_D=325MeV$}
			\end{subfigure}
			\hfill
			\begin{subfigure}{0.8\textwidth}
				\centering
				\includegraphics[width=\textwidth]{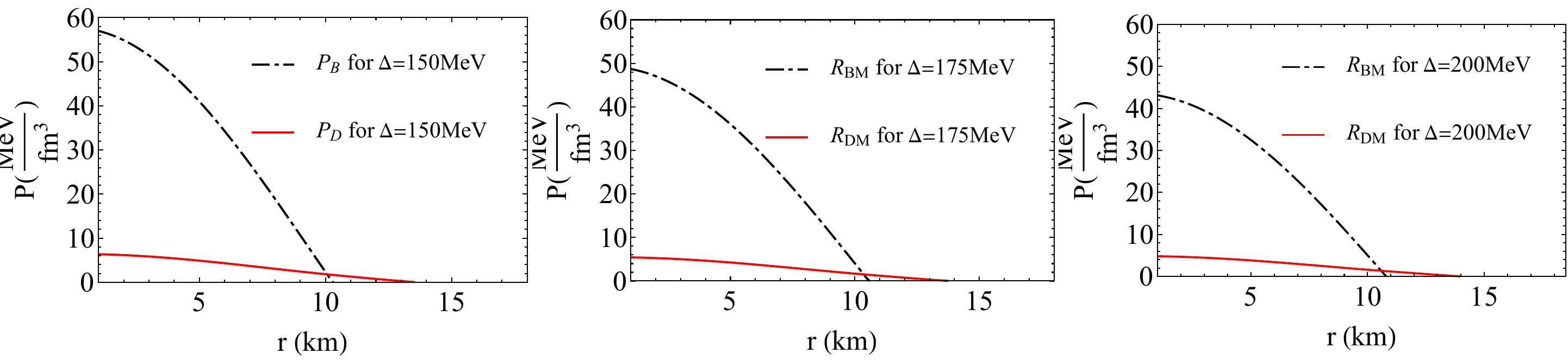}
				{$m_D=350MeV$}
			\end{subfigure}
			\caption{ $P_B(r)$ and  $P_D(r)$ corresponding to the  $M = 1.4 M_\odot$, for $fr = 10\%$. The plots include different values of $m_D$ and  $\Delta$.}
			\label{pr3}
		\end{figure*}
		Here from Fig.~\ref{mrdiagrams3} we observe that the results for $M_{\mathrm{TOV}}$ are nearly identical to those
		obtained for the same set of $m_D$ with $f_r = 5\%$. However, in contrast to the
		$f_r = 5\%$ case, we find a significant increase in $\Lambda$. A comparison of Tables~\ref{results2} and \ref{results3} clearly illustrates this
		behavior. For the same set of $m_D$, increasing $f_r$ leads to an enlargement of the
		dark matter halo, which explains why the configurations with $f_r = 10\%$  {are no longer qualitatively consistent with the observational bounds on $\Lambda$.}
		Furthermore, Fig.~\ref{tidaldiagrams3} shows that increasing $m_D$ leads to a reduction
		in $\Lambda$. This trend can be understood in terms of the associated decrease in $R_{\mathrm{out}}$, or equivalently the dark matter halo radius.
		To further clarify this point, Fig.~\ref{pr3} demonstrates that increasing $m_D$ results
		in a reduction of $R_{\mathrm{out}}$ for configurations with
		$M = 1.4\,M_\odot$. This behavior is consistent with those observed for the
		$f_r = 5\%$ configurations discussed earlier. Therefore, in the following, we consider the second set of dark matter masses with
		$f_r = 10\%$ in order to obtain configurations that  {align qualitatively with} the observational range of $70 < \Lambda_{1.4\,M_\odot} < 580$.

		\subsection{Results for $f_r = 10\%$ and Set~2 of $m_D$: $m_D = 425, 450, 475\,\text{MeV}$}
\begin{figure*}[htbp]
			\centering
			\par
			\includegraphics[width=18cm]{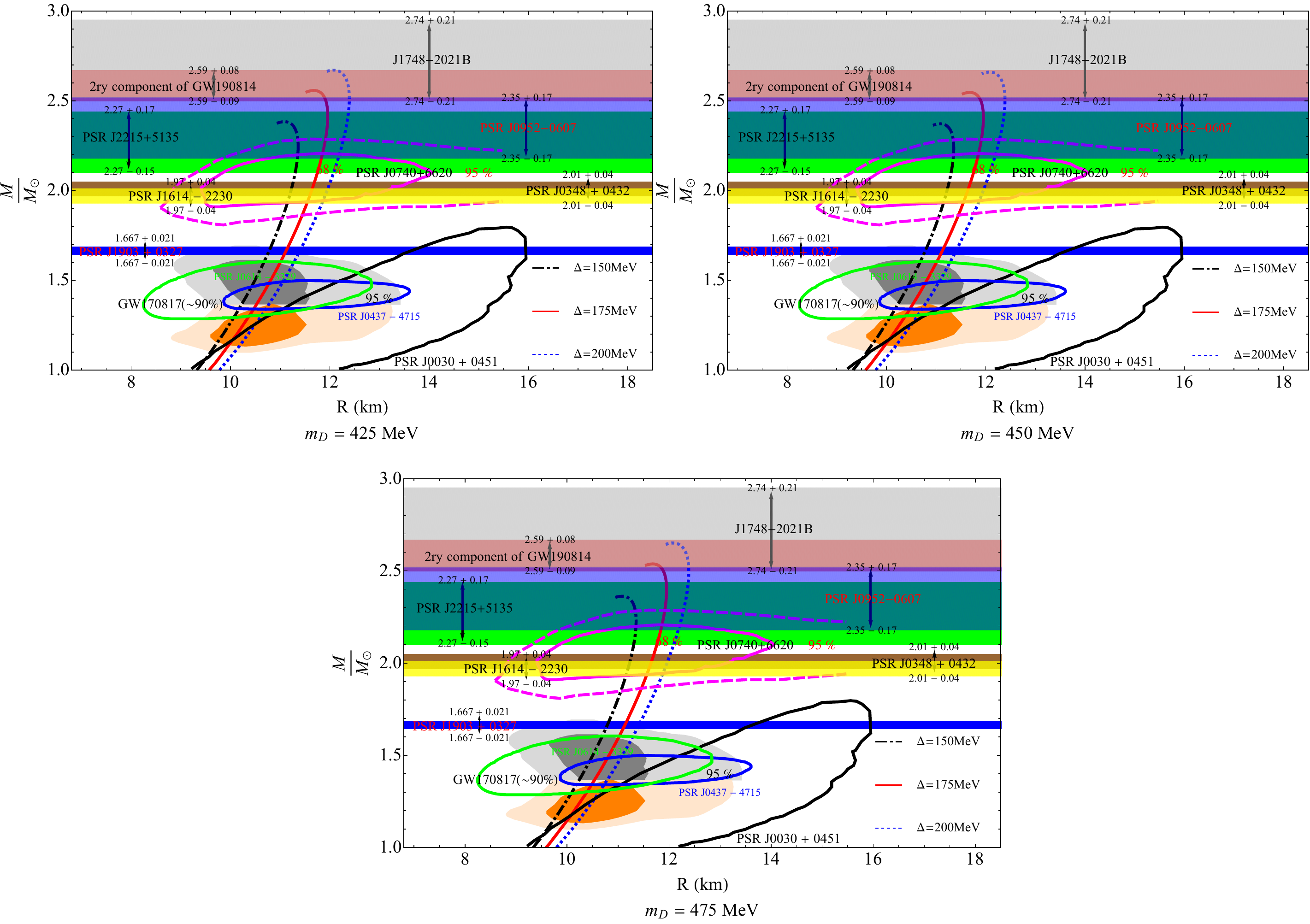}
			\caption{ {$M$--$R$ relations for $f_r = 10\%$ and different values of $m_D$ and $\Delta$. 
					The shaded horizontal bands indicate the observational mass constraints from various compact objects, while the contour regions represent mass--radius constraints inferred from astrophysical observations. 
					The radius $R$ corresponds to the SQM component radius. 
					 {The GW170817 contour is included for qualitative comparison within the present two-fluid framework.}}}
			\label{mrdiagrams4}
		\end{figure*}

\begin{figure*}[htbp]
			\centering
			\par
			\includegraphics[width=18cm]{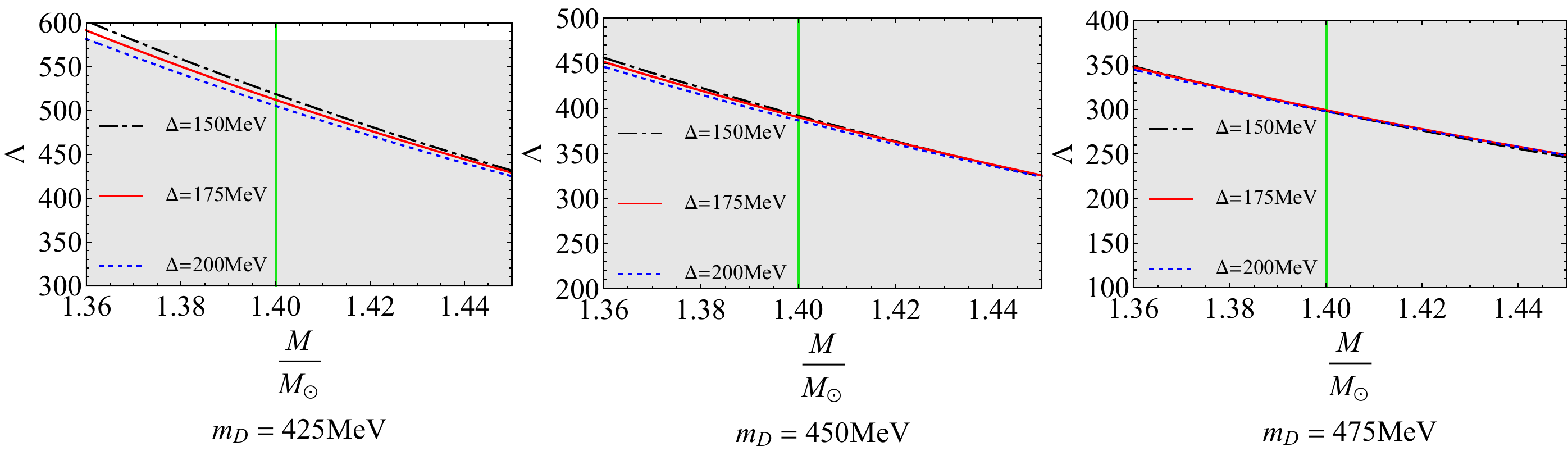}
					\caption{ {$\Lambda$ versus $M$ relations for $f_r = 10\%$ and different values of $m_D$ and $\Delta$. The shaded gray region represents the tidal-deformability range inferred from GW170817, utilized here as an  {indicative guideline} for our two-fluid model ($70 \lesssim \Lambda_{1.4M_\odot} \lesssim 580$).}}
			\label{tidaldiagrams4}
		\end{figure*}

\begin{figure*}[htbp]
			\centering
			\begin{subfigure}{0.8\textwidth}
				\centering
				\includegraphics[width=\textwidth]{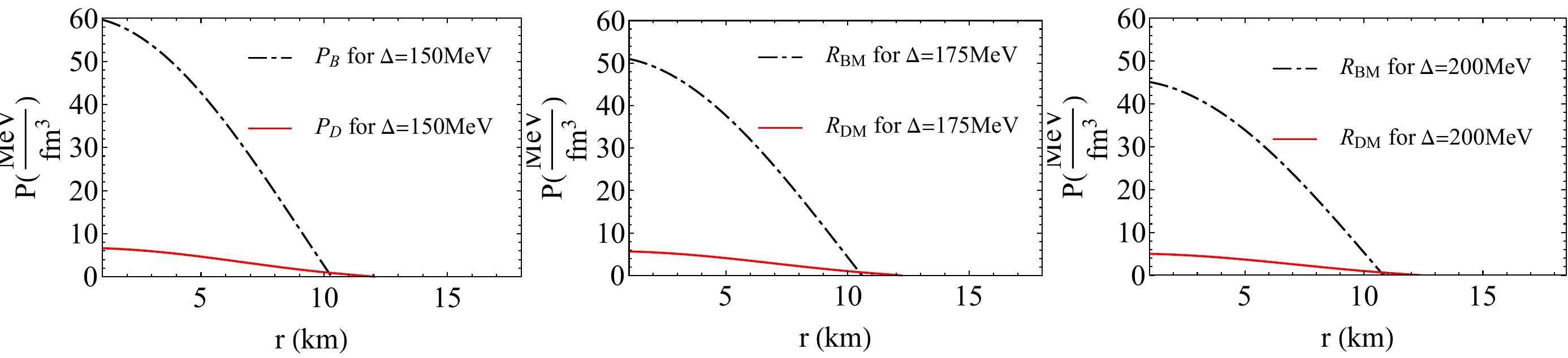}
				{$m_D=425MeV$}
			\end{subfigure}
			\hfill
			\begin{subfigure}{0.8\textwidth}
				\centering
				\includegraphics[width=\textwidth]{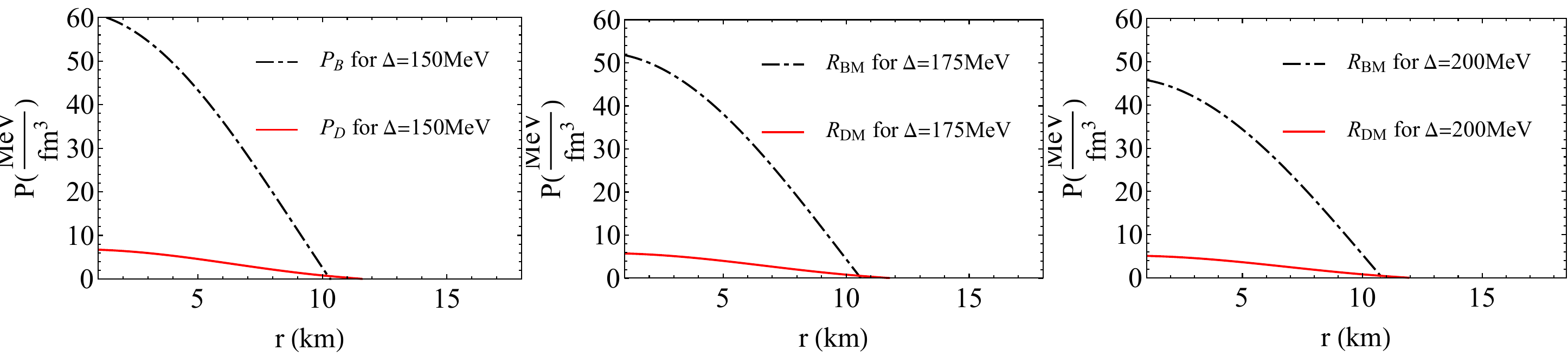}
				{$m_D=450MeV$}
			\end{subfigure}
			\hfill
			\begin{subfigure}{0.8\textwidth}
				\centering
				\includegraphics[width=\textwidth]{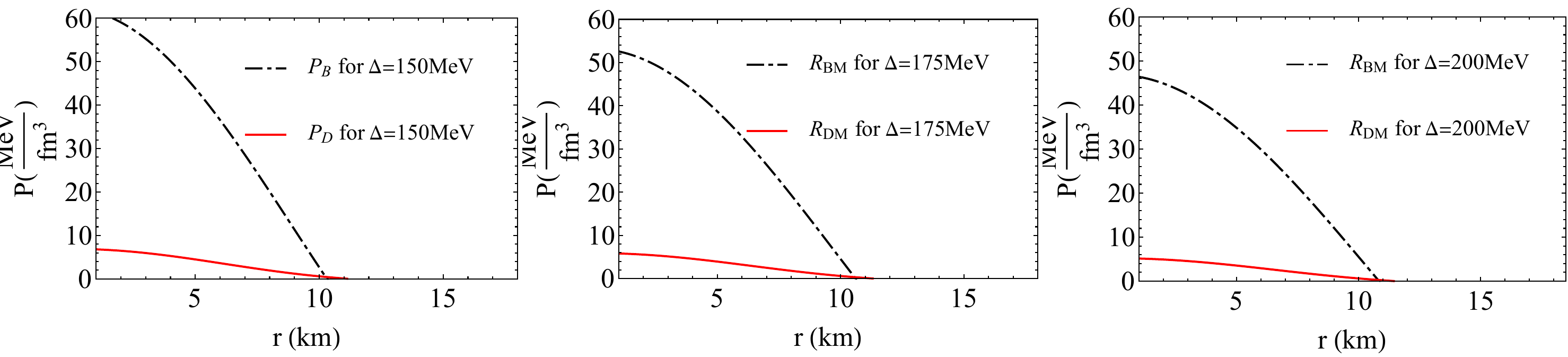}
				{$m_D=475MeV$}
			\end{subfigure}
			\caption{ $P_B(r)$ and  $P_D(r)$ corresponding to the  $M = 1.4 M_\odot$, for $fr = 10\%$. The plots include  different values of $m_D$ and  $\Delta$.}
			\label{pr4}
		\end{figure*}

		We now turn to a new parameter choice by adopting a higher dark matter mass range, $m_D = 425, 450, 475\,\mathrm{MeV}$, while fixing $f_r = 10\%$, and analyze the resulting stellar structure. As shown in Figs.~\ref{mrdiagrams4} and \ref{tidaldiagrams4}, this parameter set allows the model to accommodate massive configurations while remaining  {qualitatively consistent} with the observational bounds on $\Lambda$ (see also Table~\ref{results4}).  {The qualitative consistency}  of these results with the observational constraints can be explained in
		the same manner as in the previous cases, namely through the reduction of
		$R_{\mathrm{out}}$ and the consequent decrease in $\Lambda$. Since this mechanism has
		already been discussed in detail above, we do not repeat it here. This behavior is
		clearly illustrated in Fig.~\ref{pr4}.
			
		\begin{table}[htbp]
			\caption{		
				Structural properties of SQS for $fr=10\%$ and different values of $m_D$ and  $\Delta$. {Here, $R$ and  $R_{1.4M_\odot}$ are 
the  {SQM radius} corresponding to the $M_{TOV}$ and $M=1.4M_\odot$, respectively.}}
			\centering
			\small
			\begin{subtable}{1.0\linewidth}
				\centering
				\scalebox{0.8}{
					\begin{tabular}{|c|c|c|c|c|c|}
						\hline
						\multicolumn{6}{|c|}{ {$m_{\text{D}}$=300 MeV}} \\
						\hline
						$\Delta(MeV)$ & $\Lambda_{1.4\textup{M}_\odot}$& $R(km)$ & $M_{\text{TOV}}(\textup{M}_\odot)$ & $\frac{R_{\text{SQM}}}{R_{\text{DM}}}$ for $M=1.4\textup{M}_\odot$&${R_{1.4M_{\odot}}(km)}$\\\hline
						$150$ & 2288.56 & 11.11 & 2.47 &  0.701 & {10.29}\\ \hline
						$175$ & 2163.78 & 11.70 & 2.64 &  0.707 & {10.59} \\ \hline
						$200$ & 2088.69 & 12.09 & 2.76 &  0.711 &{10.83}\\ \hline
				\end{tabular}}
				
			\end{subtable}
			\\
			\begin{subtable}{1.0\linewidth}
				\centering
				\scalebox{0.8}{
					\begin{tabular}{|c|c|c|c|c|c|}
						\hline
						\multicolumn{6}{|c|}{ {$m_{\text{D}}$=325 MeV}} \\
						\hline
						$\Delta(MeV)$ & $\Lambda_{1.4\textup{M}_\odot}$& $R(km)$ & $M_{\text{TOV}}(\textup{M}_\odot)$ &  $\frac{R_{\text{SQM}}}{R_{\text{DM}}}$ for $M=1.4\textup{M}_\odot$&${R_{1.4M_{\odot}}(km)}$\\\hline
						$150$ & 1694.00 & 11.10 & 2.45 &  0.725 & {10.28}\\ \hline
						$175$ & 1618.18 & 11.69 & 2.62 &  0.736 & {10.58}\\ \hline
						$200$ & 1559.83 & 12.09 & 2.74 &  0.740 & {10.82}\\ \hline
				\end{tabular}}
				
			\end{subtable}
			\\
			\begin{subtable}{1.0\linewidth}
				\centering
				\scalebox{0.8}{
					\begin{tabular}{|c|c|c|c|c|c|}
						\hline
						\multicolumn{6}{|c|}{ {$m_{\text{D}}$=350 MeV}} \\
						\hline
						$\Delta(MeV)$ & $\Lambda_{1.4\textup{M}_\odot}$& $R(km)$ & $M_{\text{TOV}}(\textup{M}_\odot)$ &  $\frac{R_{\text{SQM}}}{R_{\text{DM}}}$ for $M=1.4\textup{M}_\odot$&${R_{1.4M_{\odot}}(km)}$\\\hline
						$150$ & 1256.18 & 11.08 & 2.43 &   0.757 & {10.28}\\ \hline
						$175$ & 1204.82 & 11.69 & 2.60 &  0.763 & {10.58}\\ \hline
						$200$ & 1170.72 & 12.09 & 2.72 &  0.766 & {10.82}\\ \hline
				\end{tabular}}
				
			\end{subtable}
			\label{results3}
		\end{table}
		
		
		So far, we have investigated the structural properties of SQSs in
		the CFL phase in the presence of scalar dark matter, within the framework of a
		two--fluid model. Our analysis has been carried out for configurations with different
		values of $f_r$ and $m_D$. We see that all configurations lead to the formation of a
		dark matter halo, whose size can significantly affect the value of
		$\Lambda_{1.4\,M_\odot}$. In particular, increasing $m_D$ results in a moderate decrease
		in $M_{\mathrm{TOV}}$ and a significant reduction in $\Lambda$. We have also shown that,
		for each value of $f_r$, there exist ranges of $m_D$ for which the configurations not
		only account for massive objects in the mass--gap region but also satisfy the relevant
		observational constraints.
			\begin{table}[htbp]
			\caption{		
				Structural properties of SQS for $fr=10\%$ and different values of $m_D$ and  $\Delta$. {Here, $R$ and  $R_{1.4M_\odot}$ are 
the  {SQM radius} corresponding to the $M_{TOV}$ and $M=1.4M_\odot$, respectively.}}
			\centering
			\small
			\begin{subtable}{1.0\linewidth}
				\centering
				\scalebox{0.8}{
					\begin{tabular}{|c|c|c|c|c|c|}
						\hline
						\multicolumn{6}{|c|}{ {$m_{\text{D}}$=425 MeV}} \\
						\hline
						$\Delta(MeV)$ & $\Lambda_{1.4\textup{M}_\odot}$& $R(km)$ & $M_{\text{TOV}}(\textup{M}_\odot)$ & $\frac{R_{\text{SQM}}}{R_{\text{DM}}}$ for $M=1.4\textup{M}_\odot$&${R_{1.4M_{\odot}}(km)}$\\\hline
						$150$ & 517.16 & 11.08 & 2.38 &  0.844 & {10.29}\\ \hline
						$175$ & 511.09 & 11.68 & 2.56 &  0.855 & {10.60} \\ \hline
						$200$ & 502.80 & 12.08 & 2.68 &  0.872 & {10.86}\\ \hline
				\end{tabular}}
				
			\end{subtable}
			\\
			\begin{subtable}{1.0\linewidth}
				\centering
				\scalebox{0.8}{
					\begin{tabular}{|c|c|c|c|c|c|}
						\hline
						\multicolumn{6}{|c|}{ {$m_{\text{D}}$=450 MeV}} \\
						\hline
						$\Delta(MeV)$ & $\Lambda_{1.4\textup{M}_\odot}$& $R(km)$ & $M_{\text{TOV}}(\textup{M}_\odot)$ &  $\frac{R_{\text{SQM}}}{R_{\text{DM}}}$ for $M=1.4\textup{M}_\odot$&${R_{1.4M_{\odot}}(km)}$\\\hline
						$150$ & 392.77 & 11.09 & 2.37 &  0.889 & {10.29}\\ \hline
						$175$ & 388.98 & 11.68 & 2.55 &  0.891 & {10.60}\\ \hline
						$200$ & 385.21 & 12.07 & 2.66 &  0.908 & {10.85}\\ \hline
				\end{tabular}}
				
			\end{subtable}
			\\
			\begin{subtable}{1.0\linewidth}
				\centering
				\scalebox{0.8}{
					\begin{tabular}{|c|c|c|c|c|c|}
						\hline
						\multicolumn{6}{|c|}{ {$m_{\text{D}}$=475 MeV}} \\
						\hline
						$\Delta(MeV)$ & $\Lambda_{1.4\textup{M}_\odot}$& $R(km)$ & $M_{\text{TOV}}(\textup{M}_\odot)$ &  $\frac{R_{\text{SQM}}}{R_{\text{DM}}}$ for $M=1.4\textup{M}_\odot$&${R_{1.4M_{\odot}}(km)}$\\\hline
						$150$ & 298.59 & 11.09 & 2.36 &   0.920 & {10.29}\\ \hline
						$175$ & 298.24 & 11.68 & 2.54 &  0.939 & {10.59}\\ \hline
						$200$ & 294.48 & 12.08 & 2.65 &  0.940 & {10.85}\\ \hline
				\end{tabular}}
				
			\end{subtable}
			\label{results4}
		\end{table}
		
		\section{Analysis of the Critical $m_D$ Affecting $M_{\text{TOV}}$}\label{critmD}
\begin{figure*}[htbp]
			\centering
			\par
			\includegraphics[width=18cm]{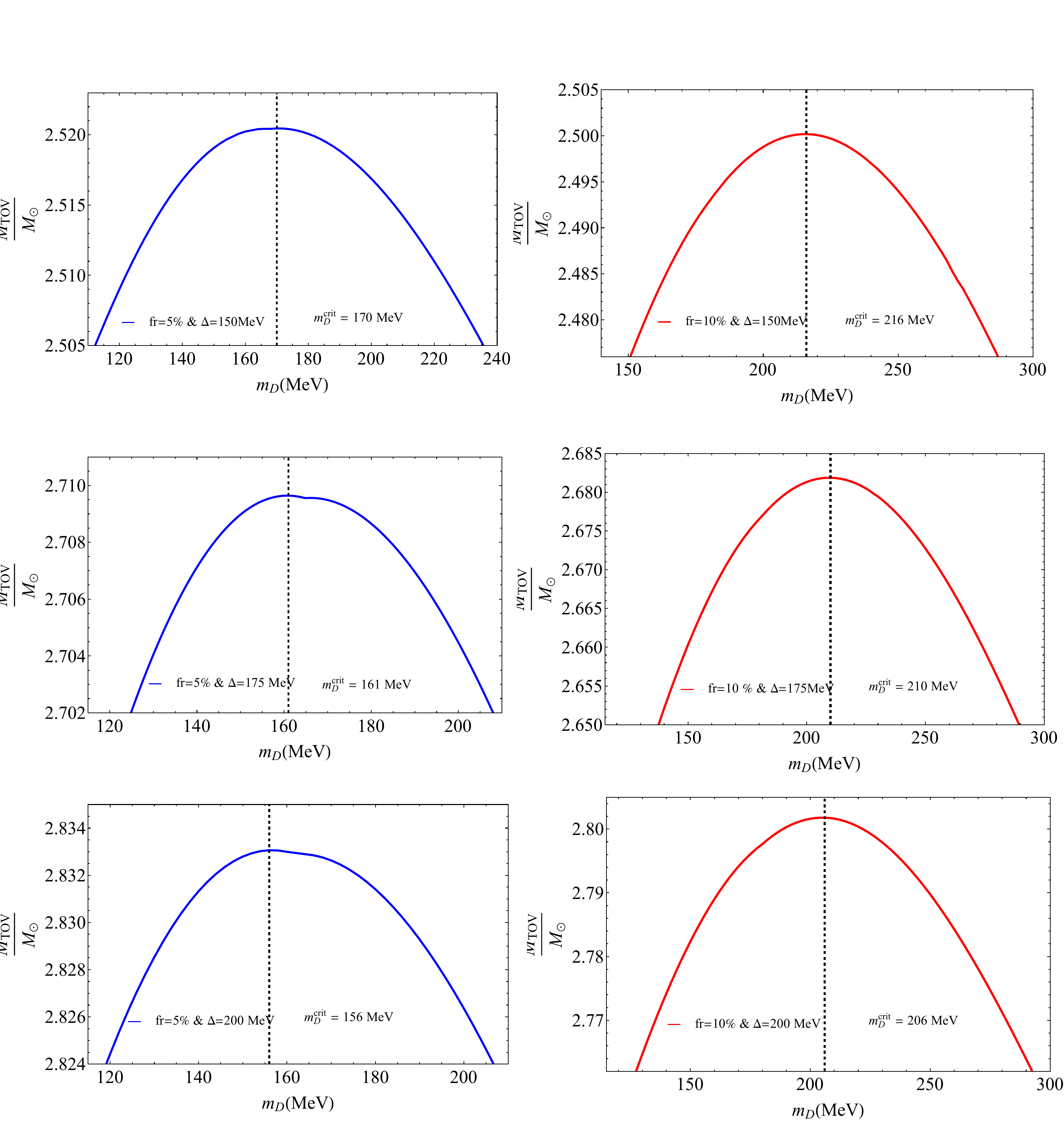}
			\caption{$M_{\text{TOV}}$ versus $m_D$ for different values of $\Delta$}
			\label{MTOVmD}
		\end{figure*}

		In the previous sections, we observed contrasting behaviors in the $M_{\text{TOV}}$ as the dark matter mass $m_D$ increased, depending on the chosen set of values. These findings suggest the existence of a non-monotonic relationship between $M_{\text{TOV}}$ and $m_D$. In this section, we investigate this behavior in more detail and show that for each fixed value of the parameter $\Delta$, there exists a critical dark matter mass beyond which further increases in $m_D$ lead to a reduction in $M_{\text{TOV}}$. Identifying this critical point provides deeper insight into how dark matter influences the stability and maximum mass of SQSs.  {It is important to note that the stability properties of the admixed star, such as $M_{\text{TOV}}$, are determined by the interplay between the SQM and dark matter components. Within our two--fluid framework, the values of $f_r$ and $m_D$ are intrinsically linked, meaning the resulting equilibrium configurations are specific to the chosen EOS and parameter set. Therefore, the following analysis presents the behavior of $M_{\text{TOV}}$ only within the configurations investigated in this study, corresponding to the representative values $f_r=5\%$ and $10\%$, and should not be interpreted as establishing a general monotonic trend over the full parameter space. Moreover, within the dark matter EOS (Eq.~(\ref{omegaBEC})), the scattering length $l_a$ directly influences the results. As mentioned, we have fixed $l_a = 1\,\text{fm}$ in this study; therefore, our findings in this section should be interpreted as being conditional on this assumed parameter space.} For further investigation of the non-monotonic dependence of the maximum mass on the dark matter mass, we examine $M_{\text{TOV}}$ as a function of $m_D$ for various values of the $\Delta$ and  $fr$. The results, shown in Fig.~\ref{MTOVmD}, reveal that for each fixed value of $\Delta$ and $fr$, there exists a critical dark matter mass, $m^{crit}_D$ at which $M_{\text{TOV}}$ reaches its maximum. Beyond this point, increasing $m_D$ leads to a decrease in $M_{\text{TOV}}$. The values of $m^{crit}_D$ for different configurations are summarized in Table~\ref{critical-md}. These results suggest that there is a threshold for the dark matter mass, beyond which its further inclusion leads to a reduction in the maximum stable mass of the star, thus placing constraints on the dark matter content compatible with stable configurations.
		\begin{table}[htbp]
			\centering
			\caption{Critical dark matter mass $m^{crit}_D$ corresponding to the maximum gravitational mass $M_{\text{TOV}}$ for various values of  $\Delta$ and  $fr$.}
			\label{critical-md}
			\begin{tabular}{|c|c|c|c|}
				\hline
				$\Delta$ (MeV) & $fr$ (\%) & $m^{crit}_D$ (MeV) & $M_{\text{TOV}}$ ($M_\odot$) \\
				\hline
				150 & 5  & 170 & 2.515 \\
				\hline
				150 & 10 & 216 & 2.505 \\
				\hline
				175 & 5  & 161 & 2.710 \\
				\hline
				175 & 10 & 210 & 2.685 \\
				\hline
				200 & 5  & 156 & 2.834 \\
				\hline
				200 & 10 & 206 & 2.800 \\
				\hline
			\end{tabular}
		\end{table}
		The results presented in Table~\ref{critical-md} and Fig.~\ref{MTOVmD} demonstrate a clear non-monotonic behavior of the maximum gravitational mass $M_{\text{TOV}}$ as a function of the dark matter mass $m_D$.  The trends observed can be summarized as follows:\\
		\textbf{I)Dependence on pairing gap $\Delta$:} \\
		For both $fr = 5\%$ and $ 10\%$, increasing $\Delta$ leads to a systematic increase in $M_{\text{TOV}}$. This behavior is consistent with the fact that a larger pairing gap enhances the stiffness of the EOS, allowing the star to support more mass. Interestingly, $m^{crit}_D$ decreases as $\Delta$ increases, implying that for stronger pairing interactions, the destabilizing effect of heavier dark matter occurs at a lower mass threshold.\\
		\textbf{II)Dependence on central dark matter pressure fraction $fr$:} \\
		For a fixed $\Delta$, increasing the $fr$ from 5\% to 10\% results in a decrease in $M_{\text{TOV}}$, indicating that the inclusion of more dark matter softens the EOS and lowers the maximum stable mass. At the same time, $m^{crit}_D$ increases with higher $fr$, meaning that configurations with more dark matter can tolerate heavier dark matter particles before reaching the turning point.\\
		\textbf{Physical interpretation:} \\
		The existence of a critical value $m^{crit}_D$ beyond which $M_{\text{TOV}}$ decreases highlights a balance between two competing effects. At low $m_D$, dark matter contributes beneficially to the pressure support and gravitational mass of the star. However, at high $m_D$, the central concentration of dark matter becomes excessive, introducing instability and reducing the star's mass-bearing capacity. 
	
Our results confirm that the interplay between the SQM EOS stiffness (regulated by $\Delta$) and the dark matter properties ($m_D$, $fr$) governs the maximum mass such stars can attain. 

		\section{summary and Conclusion}
		In this study, we investigated the structural properties of strange quark stars (SQSs) in
		the color--flavor--locked (CFL) phase admixed with scalar bosonic dark matter, modeled
		within a two--fluid formalism. The quark matter equation of state (EOS) was derived using
		perturbative QCD, incorporating the latest Particle Data Group values for the running QCD
		coupling constant and the strange quark mass, while the dark matter EOS was constructed
		based on Bose--Einstein condensation of scalar particles.
		
		To systematically assess the impact of dark matter, we explored a range of dark matter
		masses $m_D$ for different dark matter fractions in the central pressure, $f_r$, including
		$f_r = 5\%$ and $10\%$. A key parameter in our analysis is the CFL pairing gap $\Delta$,
		which quantifies the strength of quark pairing due to color superconductivity. We
		considered three representative values, $\Delta = 150$, $175$, and $200~\mathrm{MeV}$,
		corresponding to increasingly stiff EOSs, as motivated by QCD-inspired models and previous
		studies of CFL matter. Our results show that the presence of dark matter can significantly
		modify the global properties of SQSs, including their maximum mass, radius, and $\Lambda$.
		We found that, for specific ranges of the parameters $(m_D, f_r, \Delta)$, the model results  {remain qualitatively consistent} with the $\Lambda$ bounds inferred from the GW170817 event ($\Lambda_{1.4\,M_\odot} < 580$) while simultaneously accounting for several
		high-mass and ultra-massive compact objects, including
		\begin{itemize}
			\item PSR J0952--0607 ($2.35\pm0.17\,M_\odot$),
			\item PSR J0740+6620 ($2.08\pm0.07\,M_\odot$),
			\item PSR J1748--2021B ($2.74\,M_\odot$),
			\item the secondary component of GW190814 ($2.59\,M_\odot$).
		\end{itemize}
 {These results indicate that CFL quark stars admixed with dark matter can accommodate massive objects within the mass-gap region in well-defined areas of the parameter space, suggesting that the existence of a mass-gap in the $2.5$--$5\,M_\odot$ range might be less universal than previously assumed. In contrast, some pure CFL configurations, especially those producing stiffer EOSs and higher maximum masses, exhibit challenge with the GW170817 $\Lambda$ bounds, despite their stiff EOS and ability to reach large stellar masses, as demonstrated in Appendix~\ref{app:noDM_CFL}.}
		
		We also showed that a reduction of the outermost stellar radius---when determined by
		the extent of the dark matter halo---leads to a substantial decrease in the $\Lambda$ due to the strong dependence of $\Lambda$ on $R_{\mathrm{out}}$. In addition,
		the maximum gravitational mass $M_{\mathrm{TOV}}$ exhibits a non-monotonic dependence on the
		dark matter mass $m_D$, with the existence of a critical value $m_D^{\mathrm{crit}}$ for
		each choice of $f_r$ and $\Delta$, beyond which $M_{\mathrm{TOV}}$ decreases. It should be mentioned that in \cite{I.Bombaci2021} the authors suggested that the secondary component of GW190814 could be interpreted as a SQS within a single--fluid model. Although they obtained SQS configurations with \(M_{\mathrm{TOV}}\sim 2.6\,M_\odot\) and presented an interesting analysis consistent with gamma--ray burst production scenarios, their study differs from ours in several respects.	
		First, their SQM EOS is based on a purely phenomenological quark matter model presented in \cite{Alford2005}, where both the QCD coupling constant and the strange quark mass are treated as fixed parameters. Second, the authors did not examine whether their EOS satisfies the observational constraints on the $\Lambda$.
		In contrast, our two-fluid model incorporates perturbative QCD with both the running coupling constant and the running strange quark mass, and remains  {qualitatively consistent} with the observational bounds on $\Lambda$ derived from GW170817.
	
 {In conclusion, our results show that the inclusion of dark matter in quark star models can produce configurations that are  {qualitatively consistent} with mass-gap observations and $\Lambda$ bounds. The observed degeneracy between different compositions suggests that single-fluid models may not uniquely account for all astrophysical data. Consequently, incorporating two-fluid models in statistical analyses of compact star observations may be essential for a more complete and unbiased interpretation of the observational data.}
		
		
\section*{Acknowledgements}
 We wish to thank Shiraz University Research
		Council. This work is based upon research funded by Iran National
		Science Foundation (INSF) under project No. 4044248.
		
		\appendix
		
	\section{ {Structural Properties of Pure CFL SQSs with Stiff EOSs}}
		\label{app:noDM_CFL}
		 {In this appendix, we investigate the challenge between reaching mass-gap values and satisfying the $\Lambda$ constraints for pure CFL SQSs, with a specific focus on parameter sets that produce stiffer EOS.}
					
\begin{figure*}
\centering
\includegraphics[width=1\textwidth]{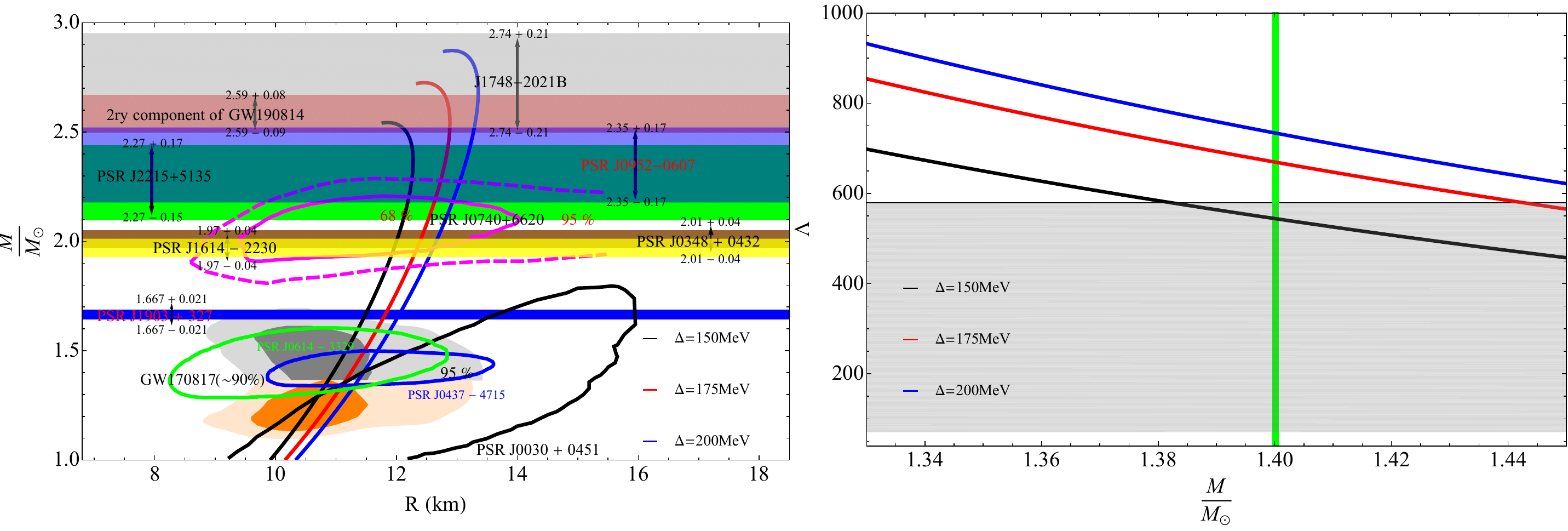}
\caption{{Left pannel: Mass--radius relation for strange quark stars composed of purely SQM in CFL phase with different values of $\Delta$.  {The shaded horizontal bands indicate the observational mass constraints from various compact objects, while the contour regions represent mass--radius constraints inferred from astrophysical observations.}  Right pannel: Dimensionless tidal deformability $\Lambda$ as a function of stellar mass. {The shaded gray region represents the tidal deformability range inferred from GW170817}}} 
\label{fig:1fluid_fig}
\end{figure*}

		The CFL phase is characterized by the formation of diquark condensations in which all three quark flavors participate symmetrically. This color superconducting phase introduces a pairing gap parameter $\Delta$, which significantly modifies the thermodynamic properties of the EOS. As $\Delta$ increases, the energy cost of breaking Cooper pairs rises, resulting in a suppression of low-energy excitations and a stiffer EOS. Physically, this increased stiffness enhances the pressure at a given energy density, allowing the star to support larger masses against gravitational collapse.
		In the main text, we examined three representative values of the pairing gap parameter: $\Delta = 150$, $175$, and $200~\mathrm{MeV}$. 
	{Here, we present the mass--radius relation and the $\Lambda$ as a function of stellar mass for pure SQS in the CFL phase, and compare these results with those obtained in the two-fluid model. Figure~\ref{fig:1fluid_fig} shows the $M-R$ relation (left panel) and the  $\Lambda$ as a function of mass (right panel) for CFL-phase SQS with different values of the gap parameter $\Delta$, in the absence of dark matter. As one can see in this figure and Table \ref{pureSQS}, these models support stellar masses exceeding $2.5M_{\odot}$, entering the observationally interesting mass-gap region. Furthermore, the results are consistent with the radius constraints inferred from GW170817. However, as illustrated in the right panel, the values of $\Lambda$ for a canonical $1.4 M_{\odot}$ star--except for $\Delta=150 MeV$-- are significantly higher than the upper limit $\Lambda_{1.4 M_{\odot}}$ inferred from GW170817. {This indicates that the $\Lambda$ constraint is restrictive within a single-fluid framework for some pure CFL parameter choices, especially those producing stiffer EOSs and higher maximum masses.} This challenge between supporting high masses and satisfying the $\Lambda$ bounds represents a fundamental limitation of the one--fluid CFL model. It strongly motivates the introducing an additional component--such as a scalar dark matter fluid, as discussed in the main text--to modify the tidal response while preserving high-mass solutions. We have found that two--fluid configurations can simultaneously satisfy both the $\Lambda$ and radius constraints. This suggests that different internal compositions--such as pure quark matter and quark matter admixed with dark matter--may lead to similar observable signatures, giving rise to degeneracies in the interpretation of gravitational-wave data. Our results for both pure SQSs and SQSs admixed with scalar dark matter are consistent with the inferred masses and radii of the GW170817 components. Moreover, the radii of pure SQSs, as well as the baryonic radii of SQSs admixed with dark matter, satisfy the observational constraint \(R_{1.4\,M_{\odot}} = 11^{+0.9}_{-0.6}\,\text{km}\) \cite{Capano} (see Tables \ref{results1}, \ref{results2}, \ref{results3}, \ref{results4}, and \ref{pureSQS}). However, the $\Lambda$ also acts as a discriminating observable, significantly reducing this degeneracy and allowing for a more clear distinction between single-fluid and two--fluid models. Therefore, incorporating two-fluid models in statistical analyses of compact star observations may be essential for a more complete and unbiased interpretation of the observational data.}
	\begin{table}
		\centering
		\caption{{Structural properties of pure SQS in CFL phase for different values of $\Delta$}}
		\begin{tabular}{|c|c|c|c|c|}
			\hline
			$\Delta(MeV)$&$R(km)$ &$\frac{M}{M_{\odot}}$&$\Lambda_{1.4\textup{M}_\odot}$& $R_{1.4\textup{M}_\odot}$ \\
			\hline
			$150$	& 11.86 & 2.54 & 544.57 & 10.94\\
			\hline
			$175$	& 12.42 & 2.73 & 670.78 & 11.25\\
			\hline
			$200$	& 12.92 & 2.87 & 731.88 & 11.44\\
			\hline
		\end{tabular}	\label{pureSQS}
	\end{table}
\section{Checking causality in our two--fluid model}
\begin{figure}
	\centering
	\par
	\includegraphics[width=8.30cm]{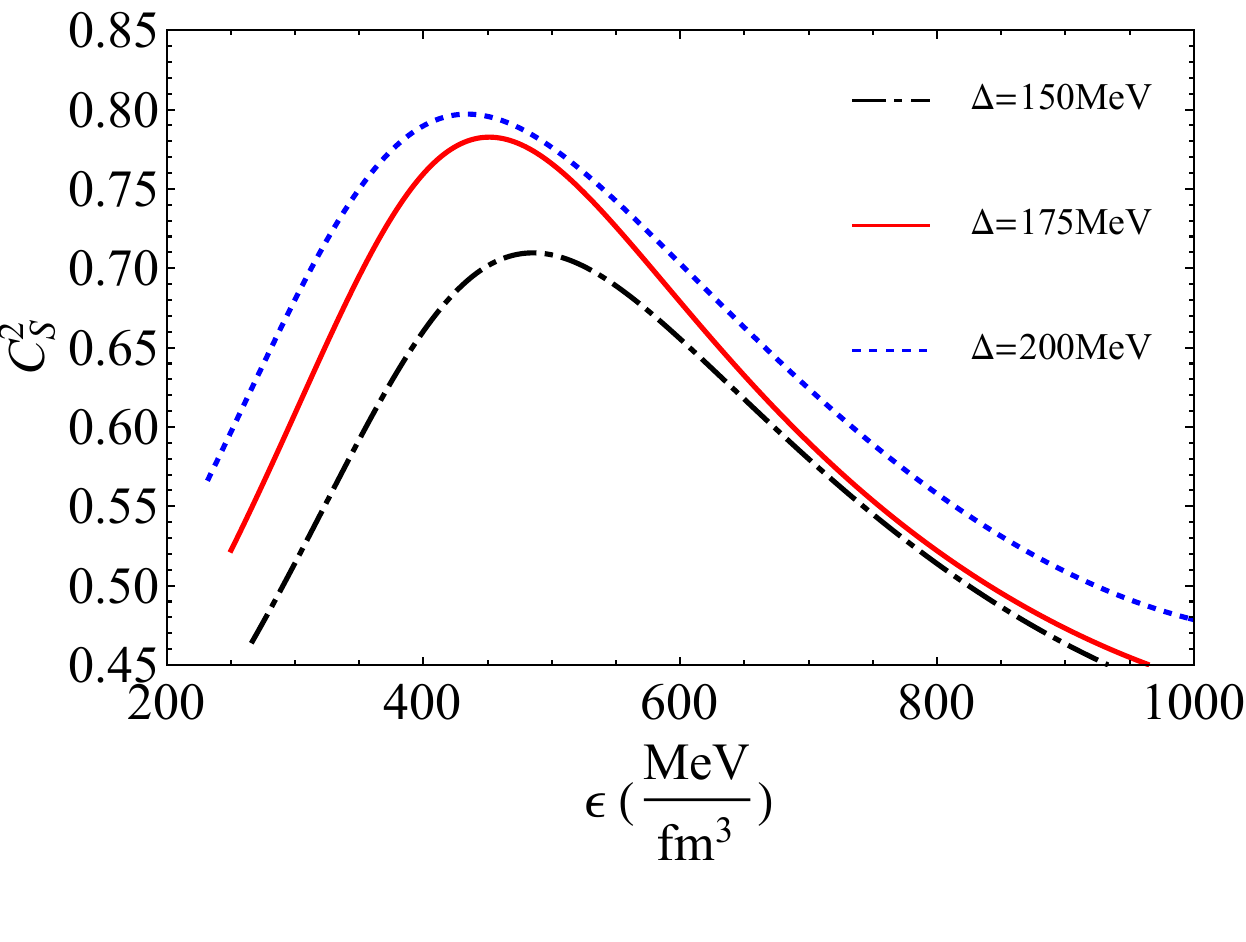}
	\caption{Speed of sound in strange quark matter versus energy density for different values of $\Delta$}
	\label{soundspeedQM}
\end{figure}

\begin{figure}
	\centering
	\par
	\includegraphics[width=8cm]{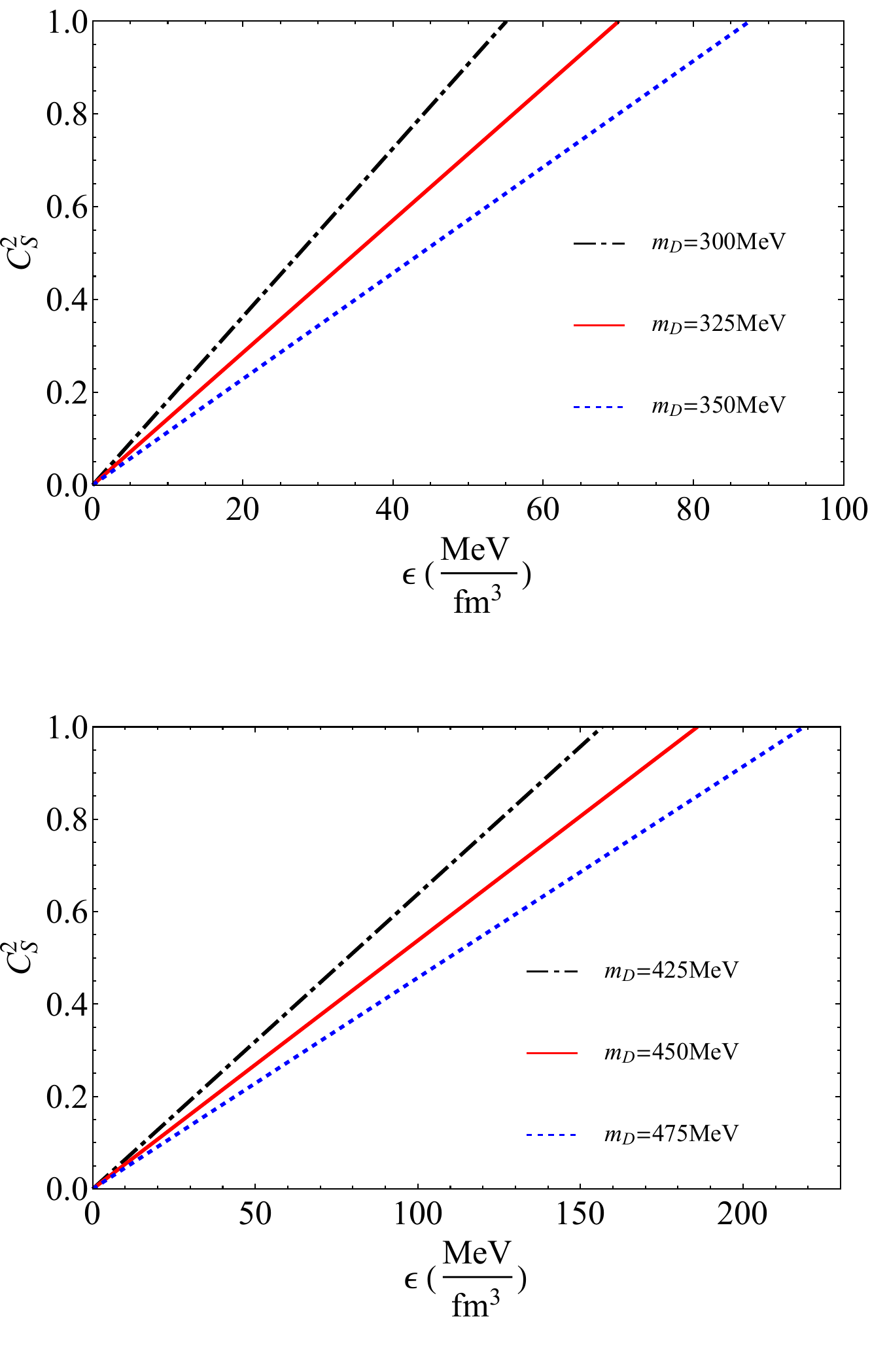}
	\caption{Speed of sound in dark matter versus energy density for different values of $m_D$}
	\label{soundspeedDM}
\end{figure}

\begin{figure}[H]
	\centering
	\par
	\includegraphics[width=8cm]{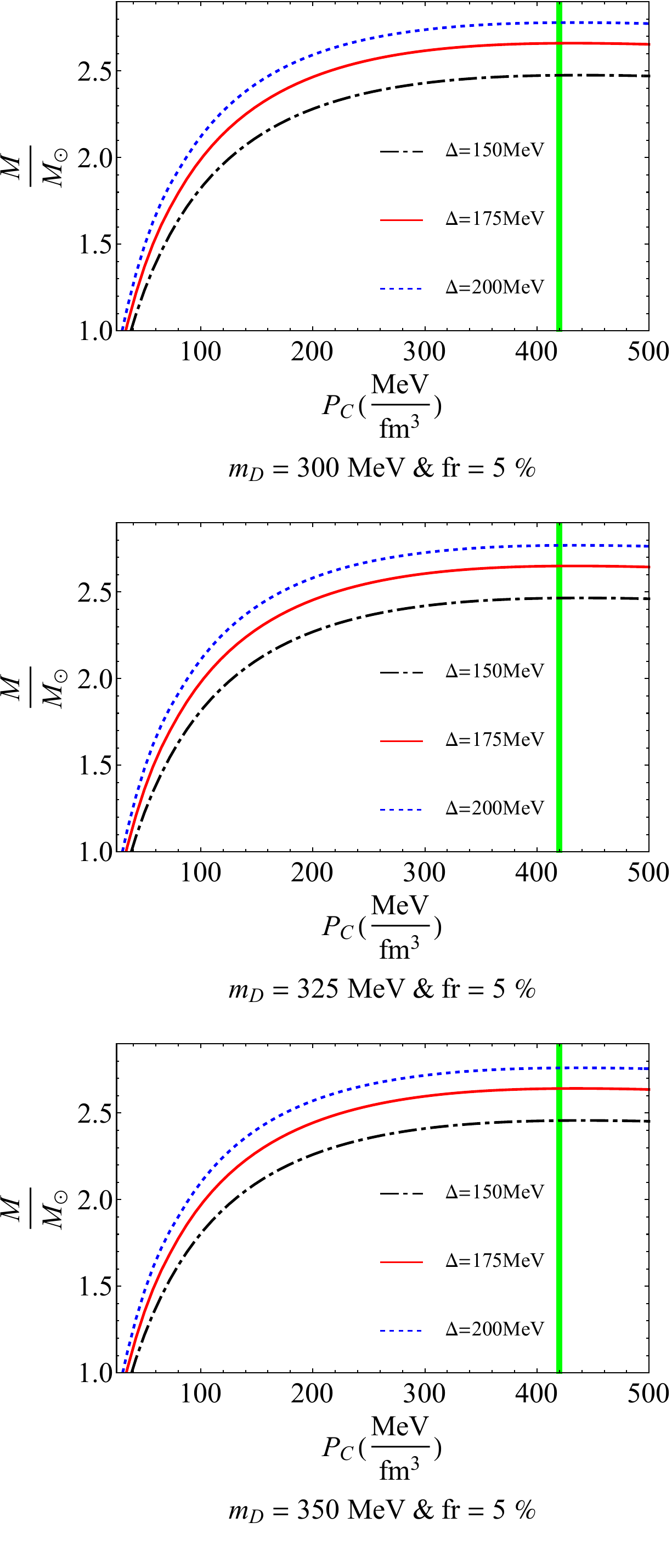}
	\caption{{Mass as a function of central pressure for \( f_r = 5\% \) at different values of \( m_D \). The vertical green line indicates the total central pressure \( P_c \approx 420\,\mathrm{MeV/fm^3} \), corresponding to \( M_{\mathrm{TOV}} \).}}
	\label{mptotal5}
\end{figure}

\begin{figure}[H]
	\centering
	\par
	\includegraphics[width=8cm]{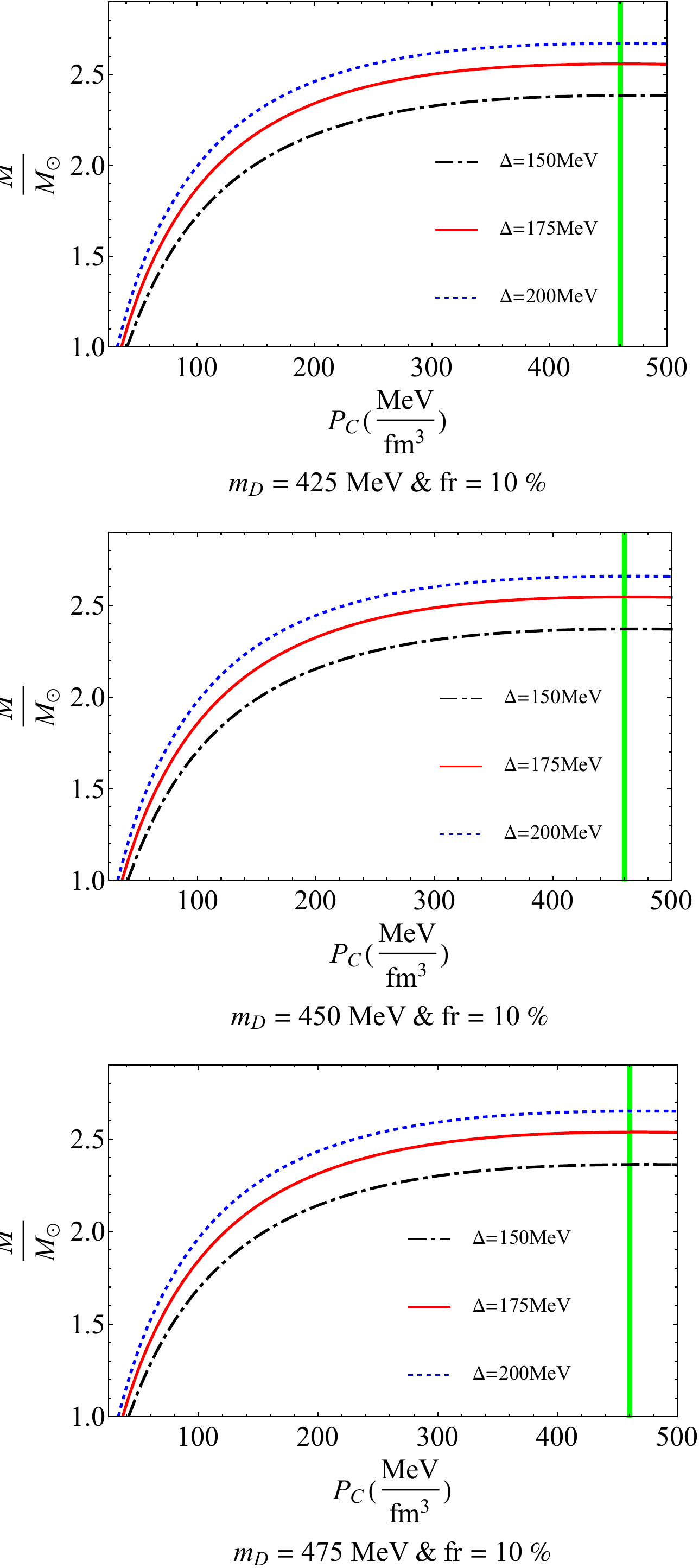}
	\caption{{Mass as a function of central pressure for \( f_r = 10\% \) at different values of \( m_D \). The vertical green line indicates the total central pressure \( P_c \approx 460\,\mathrm{MeV/fm^3} \), corresponding to \( M_{\mathrm{TOV}} \).}}
	\label{mptotal10}
\end{figure}

A natural question that arises is whether the configurations considered in our two--fluid model satisfy the causality condition. In this appendix we address this important issue. As mentioned earlier, in the two--fluid model baryonic matter and dark matter are independent
components and interact only through gravity. Consequently, two independent sound speeds exist, one for quark matter and the other for dark matter. For each fluid, the squared sound speed, \( C_S^2 = dp/d\epsilon \), must remain below unity (in natural units) in order to respect causality.
We show that this requirement is well satisfied for both the SQM fluid and the dark matter fluid.
Figure \ref{soundspeedQM} shows \( C_S^2 \) as a function of the energy density \( \epsilon \) for SQM in the CFL phase corresponding to different values of \( \Delta \). It is evident that all curves satisfy the causality condition. Moreover, increasing \( \Delta \) stiffens the EOS and consequently increases \( C_S \), which is the expected behavior in the CFL phase.
We now turn to the sound speed behavior in the second fluid, namely scalar dark matter. For brevity, we focus only on configurations  {that show qualitative consistency} with the $\Lambda$ bounds inferred from GW170817. From relation \ref{omegaBEC} it follows that \( C_S^2 \) increases linearly with the energy density.
This behavior is clearly illustrated in Fig. \ref{soundspeedDM}. As can be seen, for each value of \( m_D \) there exists a maximum energy density, \( \epsilon_{D_\text{max}} \), beyond which the causality condition is violated. Using relation \ref{omegaBEC}, one can also determine the corresponding maximum pressure of dark matter, \( P_{D_\text{max}} \), above which causality is violated. Table \ref{Pmax} lists the values of \( \epsilon_{D_\text{max}} \) and \( P_{D_\text{max}}\).
{To verify whether the configurations satisfy the causality condition in the dark matter sector, we first determine the maximum value of the total central pressure ($P_C$)  that the star can sustain before collapse. Since the maximum-mass configuration corresponds to the highest central pressure, it provides a conservative upper bound on the dark matter pressure. Therefore, if the causality condition is satisfied for this configuration, it is guaranteed to hold for all the lower pressures. To this end, we plot the stellar mass as a function of central pressure in Figs. \ref{mptotal5} and \ref{mptotal10}. The stability condition requires $\frac{\partial M}{\partial P_C}>0$ and the maximum-mass (TOV) configuration is identified by the condition $\frac{\partial M}{\partial P_C}=0$. Figures \ref{mptotal5} and \ref{mptotal10} illustrate the stellar mass versus \(P_C\) for \(f_r = 5\%\) and \(f_r = 10\%\), considering different values of \(m_D\). In each figure, the vertical green line marks the central pressure corresponding to \(M_{\mathrm{TOV}}\). These pressures are approximately \(P_c \approx 420\,\mathrm{MeV/fm^3}\) for \(f_r = 5\%\) and \(P_c \approx 460\,\mathrm{MeV/fm^3}\) for \(f_r = 10\%\). Consequently, the corresponding values of \(P_{D_\text{max}}\) are \(21\,\mathrm{MeV/fm^3}\) and \(46\,\mathrm{MeV/fm^3}\) for \(f_r = 5\%\) and \(f_r = 10\%\), respectively. Comparing these results with those listed in Table \ref{Pmax} shows that the causality condition is well respected in our configurations.} 

As shown in this appendix,  {the central dark matter pressure fraction} $f_r$ is not an unconstrained free parameter. For a given dark matter mass $m_D$, the dark matter EOS imposes a causality bound that determines the maximum admissible dark matter pressure, \( P_{D_{\text{max}}} \).
In the stellar configuration,  {the central dark matter pressure}, \(P_{D_{\text{C}}}\), is related to the total central pressure through
\[
P_{D_{\text{C}}}=f_r P_C .
\]
Therefore, causality requires
\[
P_{D_{\text{C}}}\le P_{D_{\text{max}}},
\]
which translates into an upper bound on \(f_r\) for a given central pressure.
In addition, \(f_r\) directly affects the halo structure and consequently the $\Lambda$.  {Our focus is on configurations that can remain qualitatively compatible with the $\Lambda$ bounds inferred from GW170817.} Consequently, the allowed range of \(f_r\) is jointly constrained by both the causality condition and the qualitative observational limits.
It should be emphasized that these constraints depend sensitively on \(m_D\). A value of \(f_r\) that is admissible for one dark matter mass may become disallowed for another. For example, Fig.~\ref{mptotal5} and Table~\ref{Pmax} indicate that for \(f_r=5\%\), dark matter masses below \(300\,\mathrm{MeV}\) may violate the causality condition. Moreover, as discussed in the main text, the configuration with \(m_D=300\,\mathrm{MeV}\) and \(f_r=10\%\) is  {no longer qualitatively compatible} with the $\Lambda$ range inferred from GW170817 (see Fig.~\ref{tidaldiagrams3}).  {
Therefore, determining the precise upper bound on \(f_r\) for each parameter set $\{m_D,f_r,\Delta\}$ requires a dedicated numerical exploration of the full parameter space, including repeated integrations of the coupled two-fluid TOV equations together with verification of both causality and $\Lambda$ conditions. Such a systematic parameter scan lies beyond the scope of the present work, whose primary goal is to identify representative two-fluid configurations capable of producing compact stars in the lower mass-gap region while remaining qualitatively compatible with current observational constraints.} {However, to clarify how the maximum allowed value of $f_r$ varies with $m_D$ and $\Delta$, we present the results in Table~\ref{frmax}. It should be emphasized that this table was obtained without calculating the tidal deformability. Instead, the maximum allowed values of $f_r$ were determined solely by imposing the causality condition and analyzing the mass as a function of the central pressure, following the same procedure used in Figs.~\ref{mptotal5} and \ref{mptotal5}. As shown in Table~\ref{frmax}, the maximum allowed value of $f_r$ depends on both $m_D$ and $\Delta$. However, its dependence on $m_D$ is much more pronounced, while the influence of $\Delta$ is comparatively weak.}
\begin{table}
	\centering
	\caption{Maximum allowed energy density and the corresponding pressure in scalar dark matter for different values of \( m_D \).}
	\begin{tabular}{|c|c|c|c|}
		\hline
		$f_r$ &$m_D$ ($MeV$) & $\epsilon_{max}$ ($\frac{MeV}{fm^3}$) & $P_{D_{max}}$ ($\frac{MeV}{fm^3}$) \\
		\hline
	$5\%$	&300 & 55 & 27.40\\
		\hline
	$5\%$	&325 & 70 & 34.96\\
		\hline
	$5\%$	&350 & 87 & 43.24\\
		\hline
	$10\%$	&425 & 156 & 77.65\\
		\hline
	$10\%$	&450 & 186  & 92.99\\
		\hline
	$10\%$	&475 & 217 & 107.63\\
		\hline
	\end{tabular}	\label{Pmax}
\end{table}
\begin{table*}[htbp]
	\centering
	\caption{{Maximum allowed values of $f_r$ for each $m_D$ based on the causality condition.}}
	\small
	
	\makebox[\textwidth][c]{%
		
		\begin{minipage}{0.15\textwidth}
			\centering
			\textbf{$m_{\text{D}} = 300~\mathrm{MeV}$}\\
			\vspace{2mm}
			\begin{tabular}{|c|c|c|}
				\hline
				$\Delta (MeV)$ & $f_r(max)$ \\
				\hline
				150 & 6.1\%  \\
				175 & 6.2\%  \\
				200 & 6.4\%  \\
				\hline
			\end{tabular}
		\end{minipage}
		\hspace{0.1cm}
		
		\begin{minipage}{0.15\textwidth}
			\centering
			\textbf{$m_{\text{D}} = 325~\mathrm{MeV}$}\\
			\vspace{2mm}
			\begin{tabular}{|c|c|c|}
				\hline
				$\Delta (MeV)$ & $f_r(max)$ \\
				\hline
				150 & 7\% \\
				175 & 7.8\%  \\
				200 & 8\%  \\
				\hline
			\end{tabular}
		\end{minipage}
		\hspace{0.1cm}
		
		\begin{minipage}{0.15\textwidth}
			\centering
			\textbf{$m_{\text{D}} = 350~\mathrm{MeV}$}\\
			\vspace{2mm}
			\begin{tabular}{|c|c|c|}
				\hline
				$\Delta (MeV)$ & $f_r(max)$ \\
				\hline
				150 & 9.1\%  \\
				175 & 9.4\%  \\
				200 & 9.6\%  \\
				\hline
			\end{tabular}
		\end{minipage}
		\hspace{0.1cm}
		
		\begin{minipage}{0.15\textwidth}
			\centering
			\textbf{$m_{\text{D}} = 425~\mathrm{MeV}$}\\
			\vspace{2mm}
			\begin{tabular}{|c|c|c|}
				\hline
				$\Delta (MeV)$ & $f_r(max)$ \\
				\hline
				150 & 15\%  \\
				175 & 15.8\%  \\
				200 & 16\%  \\
				\hline
			\end{tabular}
		\end{minipage}
		\hspace{0.1cm}
		
		\begin{minipage}{0.15\textwidth}
			\centering
			\textbf{$m_{\text{D}} = 450~\mathrm{MeV}$}\\
			\vspace{2mm}
			\begin{tabular}{|c|c|c|}
				\hline
				$\Delta (MeV)$ & $f_r(max)$ \\
				\hline
				150 & 18\% \\
				175 & 18.4\%  \\
				200 & 19\%  \\
				\hline
			\end{tabular}
		\end{minipage}
		\hspace{0.1cm}
		
		\begin{minipage}{0.15\textwidth}
			\centering
			\textbf{$m_{\text{D}} = 475~\mathrm{MeV}$}\\
			\vspace{2mm}
			\begin{tabular}{|c|c|c|}
				\hline
				$\Delta (MeV)$ & $f_r(max)$ \\
				\hline
				150 & 20.2\%  \\
				175 & 20.5\%  \\
				200 & 20.8\%  \\
				\hline
			\end{tabular}
		\end{minipage}
		
	}
	
	\label{frmax}
\end{table*}

	\end{document}